\documentclass[10pt,twocolumn,letterpaper]{article}

\usepackage{cvpr}
\usepackage{times}
\usepackage{epsfig}
\usepackage{graphicx}
\usepackage{amsmath}
\usepackage{amssymb}
\usepackage{amsthm}

\usepackage[pagebackref=true,breaklinks=true,letterpaper=true,colorlinks,bookmarks=false]{hyperref}

\usepackage{epsfig}\graphicspath{{./images/},{./images/general/},{./images/randgraph/},{./images/hepth/}}
\usepackage{bbm}\usepackage{caption}\usepackage{subcaption}%\usepackage{algpseudocode}
\usepackage[linesnumbered,ruled]{algorithm2e}

\SetCommentSty{mycommfont}
\usepackage{url}
\usepackage{color}
\cvprfinalcopy % *** Uncomment this line for the final submission

\def\cvprPaperID{3394} % *** Enter the CVPR Paper ID here

\ifcvprfinal\fi

\newcommand{\mb}{\mathbf}

\newtheorem{mythm}{Theorem}

\newtheorem{mydef}{Definition}

\newcommand{\denselist}{\itemsep 0pt\parsep=1pt\partopsep 0pt}
\newcommand{\bitem}{\begin{itemize}\denselist}
\newcommand{\eitem}{\end{itemize}}
\newcommand{\benum}{\begin{enumerate}\denselist}
\newcommand{\eenum}{\end{enumerate}}

\def\cl{{\cal L}}
\def\ch{{\cal H}}
\def\ck{{\cal K}}
\def\cg{{\cal G}}
\def\ce{{\cal E}}
\def\cv{{\cal V}}
\def\cc{{\cal C}}

\def\cp{{\cal P}}

\def\st{\textnormal{s.t.}}

\def\mb{\mathbf}
\def\mbx{\mb{X}}
\def\bbx{\overline{\mbx}}
\def\mbw{\mb{W}}
\def\mba{\mb{A}}
\def\mbb{\mb{B}}
\def\mbe{\mb{E}}
\def\mby{\mb{Y}}
\def\mbz{\mb{Z}}

\def\mbm{\mb{M}}
\def\mbi{\mathbf{I}}
\def\tn{\textnormal}

\def\tcb{\textcolor{black}}

\begin{document}

%%%%%%%%% TITLE
\title{Distributable Consistent Multi-Object Matching}

\author{Nan Hu\\
Stanford University\\
%nanhu@stanford.edu\\
\and
Qixing Huang\\
UT Austin\\
%huangqx@cs.utexas.edu\\
\and
Boris Thibert\\
UG Alpes\\
%Boris.Thibert@univ-grenoble-alpes.fr\\
\and
Leonidas Guibas\\
Stanford University\\
%guibas@cs.stanford.edu\\
}

%\author{Nan Hu \hspace{20pt} Qixing Huang \hspace{20pt} Boris Thibert \hspace{20pt} Leonidas Guibas\\
% Stanford University\\
% Stanford, CA, USA\\
% \texttt{\small pkuhunan@gmail.com, huangqx@cs.utexas.edu, boris.thibert@imag.fr, guibas@cs.stanford.edu}{\small {}}
% \and

\maketitle
%\thispagestyle{empty}

%%%%%%%%% ABSTRACT
\begin{abstract}
In this paper we propose an optimization-based framework to multiple object matching. The framework takes maps computed between pairs of objects as input, and outputs maps that are consistent among all pairs of objects. The central idea of our approach is to divide the input object collection into overlapping sub-collections and enforce map consistency among each sub-collection. This leads to a distributed formulation, which is scalable to large-scale datasets. We also present an equivalence condition between this decoupled scheme and the original scheme. Experiments on both synthetic and real-world datasets show that our framework is competitive against state-of-the-art multi-object matching techniques.
\end{abstract}

%%%%%%%%% BODY TEXT
\section{Introduction}

Object matching techniques have been widely used in many fields of computer vision, including 2D and 3D image analysis, object recognition, biomedical identification, and object tracking. There is a rich literature on finding meaningful approximate isomorphisms between pair of objects that are represented as graphs~\cite{umeyama88,cour2006,emms2009,gori2005,wilson2008,egozi2012,Hu2013,Hu2014}. Many tasks, however, require to solve the so-called multi-object matching problem, i.e., finding consistent maps among all pairs of objects within a collection. Examples include non-rigid structure from motion \cite{Bregler2000, Dai2012} and shared object discovery \cite{ChenXL2014}. In this context, a central task is how to utilize the data collection as a regularizer to improve the maps computed between pairs of objects in isolation~\cite{Huang2013,Chen2014, Zhou2015}.

A generic constraint that one can utilize to improve maps among a collection is the so-called \emph{cycle consistency} constraint, namely composition of maps along any two paths sharing the same starting and end objects are identical. A technical challenge of utilizing this constraint is that it is impossible to check all cycles for consistency, due to the fact that the number of paths increase exponentially with the total number of objects. Recent works on joint matching have shown that the cycle consistency constraint can be translated into a much more manageable constraint, i.e., the data matrix that stores pair-wise maps in blocks is positive semidefinite and low-rank \cite{Huang2013,Pachauri2013}. Based on this connection, people have formulated multi-object matching as solving semidefinite programs (or SDP), which are convex relaxations of the corresponding matrix recovery problem. These algorithms achieved near-optimal exact recovery conditions~\cite{Huang2013, Chen2014}. On the other hand, solving semidefinite programs are computationally expensive. In a recent work, Zhou et al~\cite{Zhou2015} attempt to address the computational issue using alternating minimization for efficient low-rank matrix recovery.

In this paper, we propose a novel framework that utilizes the cycle-consistency constraint in a hierarchical manner for scalable multiple object matching. We show how to apply this framework to extend the methods described in~\cite{Huang2013,Chen2014,Zhou2015}.
%The key idea is to relax the cycle-consistency constraint properly, avoiding consider the problem of relaxing the global cycle consistent constraint,  
%From a practical point of view, e.g. in category-specific multiple object matching, chances are not all objects are visually similar to each other to have a meaningful pairwise matching. For example, to match different styles of chairs, a chair with no arm might be more similar to a chair with low arms than a chair with high arms.
In particular, instead of jointly imposing the global consistency constraint among all pair-wise maps~\cite{Huang2013, Chen2014, Zhou2015}, we split the input object collection into overlapping subsets, and impose consistency within each subset. We then impose consistency between maps across the subsets. Interestingly, we show that by combing these two consistency constraints together, we can guarantee global consistency under mild conditions (See Section \ref{sec:consistency}). Yet computationally, such a decoupled approach yields significant performance gains, when compared with existing approaches.

\subsection{Related Work}

Early works on multi-object matching (e.g.,~\cite{zass2008, Lee2011}) extend pairwise matching schemes to the multi-object setting without explicitly considering the map consistency constraint. \cite{Zach2010, Nguyen2011} proposed to detect inconsistent cycles, and formulate multi-object matching as solving combinatorial optimizations, i.e., removing bad maps to break all inconsistent cycles. Recently, people have proposed to formulate non-convex optimization problems by using the cycle
consistency constraint as an explicit constraint for either pixel-wise flow computation \cite{Zhou2015}, sparse feature matching \cite{Yan2015}, sparse shape modeling \cite{Cosmo2016}, or structure from motion \cite{Pachauri2014}. These problems are, as a consequence, hard to solve and do not admit exact recovery conditions. Recent works
\cite{Huang2013, Pachauri2013} showed that consistent maps could be extracted
from the spectrum of a data matrix that encodes pair-wise maps in blocks. Along this line of research, Huang and Guibas \cite{Huang2013} proposed an elegant solution by formulating the problem as convex relaxation and discussed the theoretical conditions for exact recovery. The result is further analyzed in \cite{Chen2014} under the condition that the underlying rank of the variable matrix is known or can be reliably estimated. Yan et al.~\cite{Yan_ICCV2015, Yan_TIP2015} also proposed matrix factorization based methods to enforce the cycle-consistency constraint. These methods, however, are not scalable to large-scale datasets, due to the cost of solving semidefinite programs. Zhou et al.~\cite{Zhou2015} enforce the positive semidefinite constraint using explicit low-rank factorizations, leading to improved computational efficiency. In contrast to these methods, our approach opens a new direction to  enforcing the cycle-consistency constraint, i.e., by splitting the datasets into overlapping subsets. This leads to further improvements in terms of computational efficiency. Most recently, Leonardos et. al. \cite{Leonardox2017} proposed a distributed consensus-based algorithm as an extension of \cite{Pachauri2013}. Their method, however, cannot handle partial matches.

%Namely, we are showing that under some well-defined conditions, local consistency implies global consistency. Simple positive and negative examples is shown in Figure \ref{fig:cover_examples}.

We organize the reminder of this paper as follows. First, we discuss the problem setup and analyze the conditions in Section~\ref{sec:consistency}. Second, we discuss the formulation of our approach in Section~\ref{sec:formulation}. In Section~\ref{sec:solver}, we present an alternating direction method of multipliers (ADMM) for solving the induced optimization problem, leading to a parallel algorithm via generalized message passing. Last but not the least, we demonstrate the effectiveness of our approach on both synthetic and real examples in Section~\ref{sec:exp}.

%-------------------------------------------------------------------------

\section{Consistency}
\label{sec:consistency}

In this section, we extend the cycle-consistency formulation described in~\cite{Huang2013} to the distributed setting. The key result is a sufficient condition on which cycle-consistency among sub-collections induces global cycle-consistency.

We begin with introducing the notations that are necessary to formally state this sufficient condition. For simplicity, we assume maps between objects are given by permutations. However, the argument can be easily extended to the case where objects are partially similar with each other. Formally speaking, we consider a map graph $\cg = (\cv = (H_1, \cdots, H_{n}), \ce)$. The vertex set $\cv$ consists of objects to be matched, and each object $H_i$ is given by $m$ points (e.g., key points extracted from an image). The edge set $\ce$ connects a subset of pairs of objects. Each edge $(i,j)\in\ce$ is associated with a permutation $\phi_{ij}: H_i \rightarrow H_j$. We first define the global consistency of $\phi_{ij}, \forall (i,j)\in \ce$:

\begin{mydef}[Cycle Consistency]
A map graph $\cg=\{\cv,\ce\}$ is \emph{cycle consistent} if for every node $v_i$ and every cycle $v_i-v_{i_1}-\cdots-v_{i_k}-v_i$, the composite map along this cycle is the identity map, i.e.
\[
\phi_{ii_1}\circ\cdots\circ\phi_{i_k i} = \emph{identity}.
\]
\end{mydef}

Now we introduce an equivalent formulation of enforcing the cycle-consistency among $\cg$ as enforcing the cycle-consistency among subgraphs of $\cg$, if these subgraphs satisfy certain conditions. Towards this end, we introduce two conditions among collection of subgraphs of $\cg$. The first condition concerns a pair of sub-graphs:

\begin{mydef}[Joint Normal]
Let $\cg_i=\{\cv_i,\ce_i\}$,  $\cg_j=\{\cv_j,\ce_j\}$ be the two subgraphs of $\cg = \{\cv, \ce\}$. We say $\cg_i$ and $\cg_j$ are \textsl{joint normal} if the vertex sets $\cv_i \setminus \cv_j$ and $\cv_j \setminus \cv_i$ are not connected by any edge of $\ce$:
\[
(s,t)\notin\ce, ~\forall H_s\in\cv_i\backslash\cv_j, H_t\in\cv_j\backslash\cv_i
\]
\end{mydef}

\begin{figure}
\begin{subfigure}{.48\linewidth} \centering \includegraphics[width=0.96\linewidth]{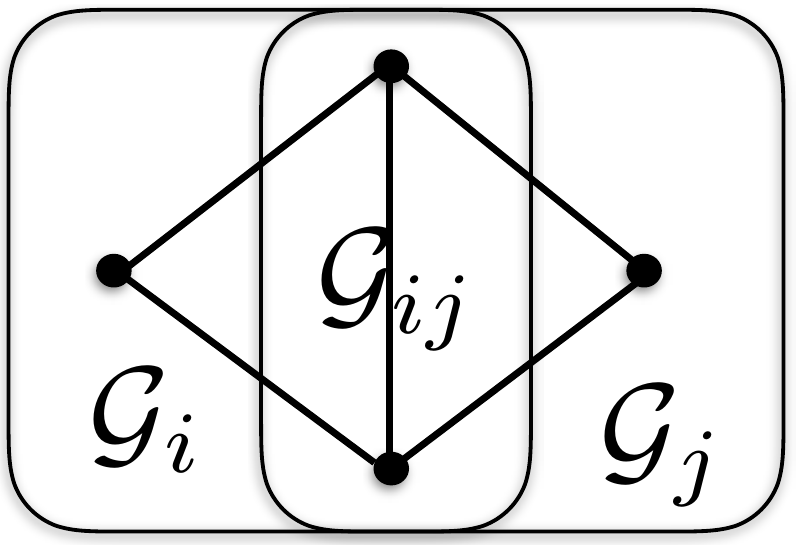}
\caption{{\it joint normal}}
\end{subfigure}
\begin{subfigure}{.48\linewidth} \centering \includegraphics[width=0.96\linewidth]{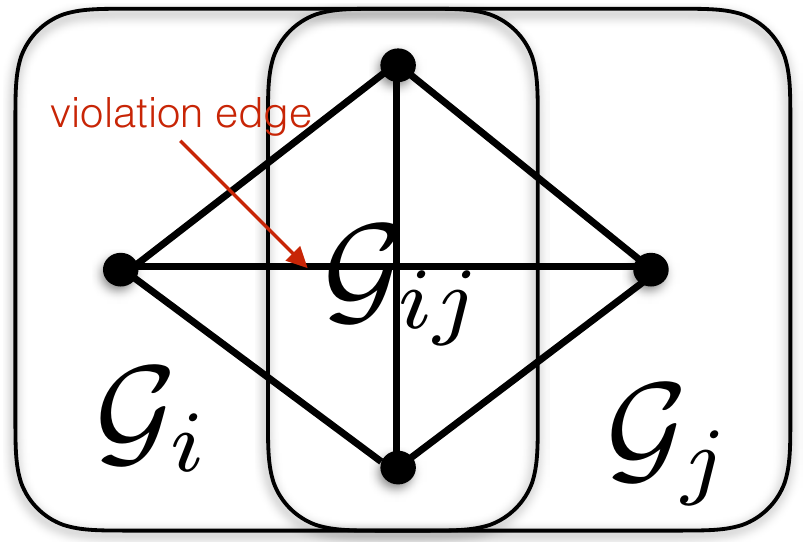}
\caption{Counter {\it joint normal}}
\end{subfigure}
\caption{Examples of subgraphs that are (a) joint normal and (b) not joint normal.}
\label{fig:joint_normal}
\end{figure}
As illustrated in Figure~\ref{fig:joint_normal}, two subgraphs $\cg_i=\{\cv_i,\ce_i\}$ and $\cg_j=\{\cv_j,\ce_j\}$ are {\it joint normal} if their common subgraph $\{\cv_i\cap \cv_j,\ce_i \cap \ce_j\}$ is either (i) empty, or (ii) connected, and there is no edge between a vertex in one subgraph to a vertex in the other subgraph except those in the common subgraph. In contrast, the two subgraphs illustrated in Figure~\ref{fig:joint_normal}(b) are not joint normal since there exists an edge that connects the non-overlapping sets of these two subgraphs.

The second condition concerns a topological constraint among all the sub-graphs. We state this condition using the notation of simplicial complex as detailed below:
\begin{mydef}[Cover Complex]
%Let the set $\{\cv_i,\ce_i\}$ be a cover of $\cg$, i.e. $\cup_i\cv_i = \cv$ and $\cup_i\ce_i=\ce$, where $\ce_i$ is the set of edges induced by $\cv_i$ from $\cg$. We call the graph $\cc_\cg=\{\cv_\cg,\ce_\cg\}$ the cover graph of $\cg$ if $\cv_\cg=\{\cv_i\}$ and $(i,j)\in\ce_\cg$ if $\cv_i\cap\cv_j\neq\varnothing$.
Let $\{\cg_i = (\cv_i, \ce_i), 1\leq i \leq K\}$ be a set of sub-graphs that cover $\cg$, i.e., $\cup_{i=1}^{K}\cv_i = \cv$. We define the cover complex $\ck$ of these sub-graphs $\{\cg_i = (\cv_i, \ce_i), 1\leq i \leq K\}$ so that $\ck$ collects every subset $ \{i_1, \cdots, i_{k}\}\subset \{1, \cdots, K\}$ if the intersections of $\cv_{i_1}, \cdots, \cv_{i_k}$ is non-empty, i.e. $\cap_j\cv_{i_j}\neq\varnothing$.
\end{mydef}
% In addition, an edge $(i,j)\in\ce_\cg$ of the cover graph $\cc_\cg$ is said to be {\it connect in $\cg$} if $\cv_i\cap\cv_j$ is connected in $\cg$ and the cover graph $\cc_\cg$ is {\it connect in $\cg$} if every edge of $\cc_\cg$ is connected in $\cg$.
 
%Note that $\ck$ is not necessarily triangulated. Equipped with the prerequisite, we have our main theorem which provides necessary and sufficient condition on which the multi-graph matching is globally cycle consistent.
We now state the decoupled cycle-consistency theorem that relates the global cycle consistency and the cycle consistency on each subgraph:
 
\begin{mythm}[Decoupled Cycle-Consistency]
\label{thm:consistency}
Let $\cg$ be a map graph, $\cg_1, \cdots, \cg_{K}$ be a cover of $\cg$, and $\ck$ be the cover complex. Then $\cg$ is cycle consistent if
\benum
\item $\cg_i$ is \emph{cycle consistent} $\forall i$,
\item $\cg_i$ and $\cg_j$ are \emph{joint normal} $\forall (i,j)\in\ce$,
\item $\ck$ is \emph{simply connected} (c.f. \cite{Hatcher2002}).
\eenum
\end{mythm}

Here we say the cover complex $\ck$ is \emph{simply connected} if every closed curve can be deformed to a point (or in other words the region enclosed by this curve has no-holes). Please refer to~\cite{Hatcher2002} for a more general definition. This theorem states that the cycle consistency property on each sub-graph would be propagated to the global consistency, if the cover complex $\ck$ satisfies the conditions stated in Theorem~\ref{thm:consistency}. %if the cover complex $\ck$ is connected, then the global consistency is obtained from the cycle consistency among each sub-graph, given that each pair of sub-graphs are joint normal.
The proof to Theorem \ref{thm:consistency} is left to the supplementary material.

\begin{figure}
\begin{subfigure}{.96\linewidth}
\centering
\includegraphics[height=3cm]{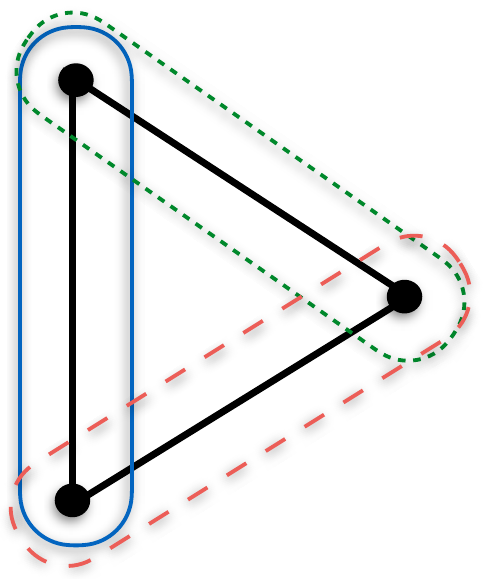}
\hspace{20pt}
\includegraphics[height=3cm]{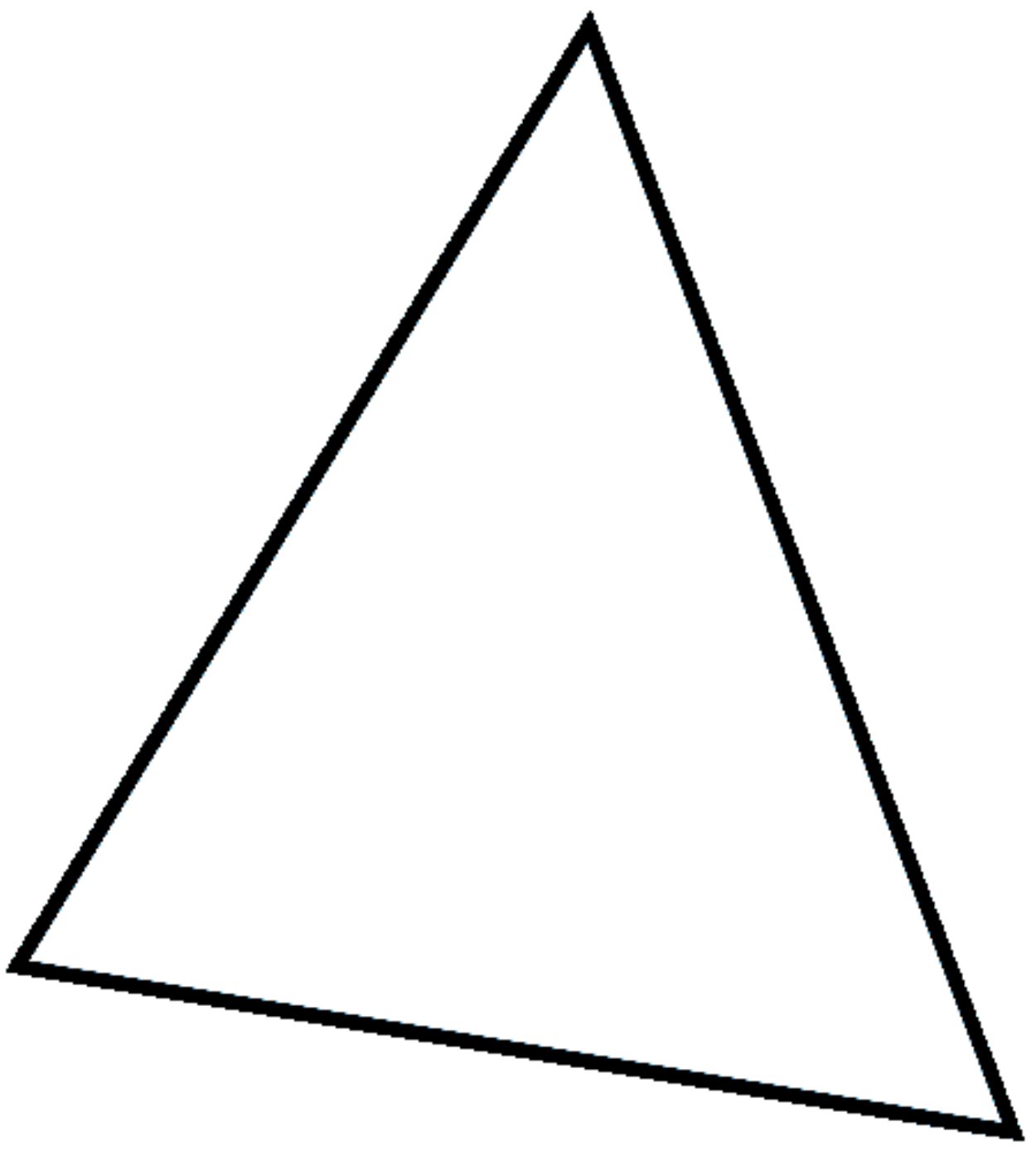}
\caption{Subgraphs that forms an empty triangle cover complex}
\end{subfigure}
\begin{subfigure}{.96\linewidth}
\centering
\includegraphics[height=3cm]{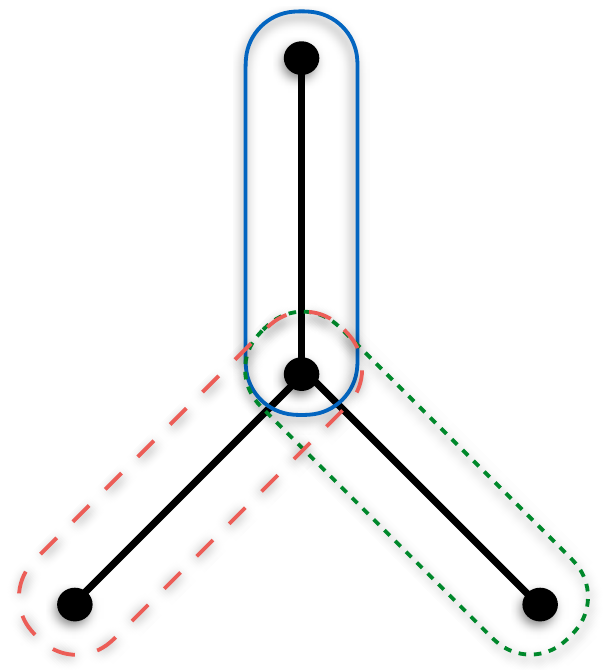}
\hspace{20pt}
\centering \includegraphics[height=3cm]{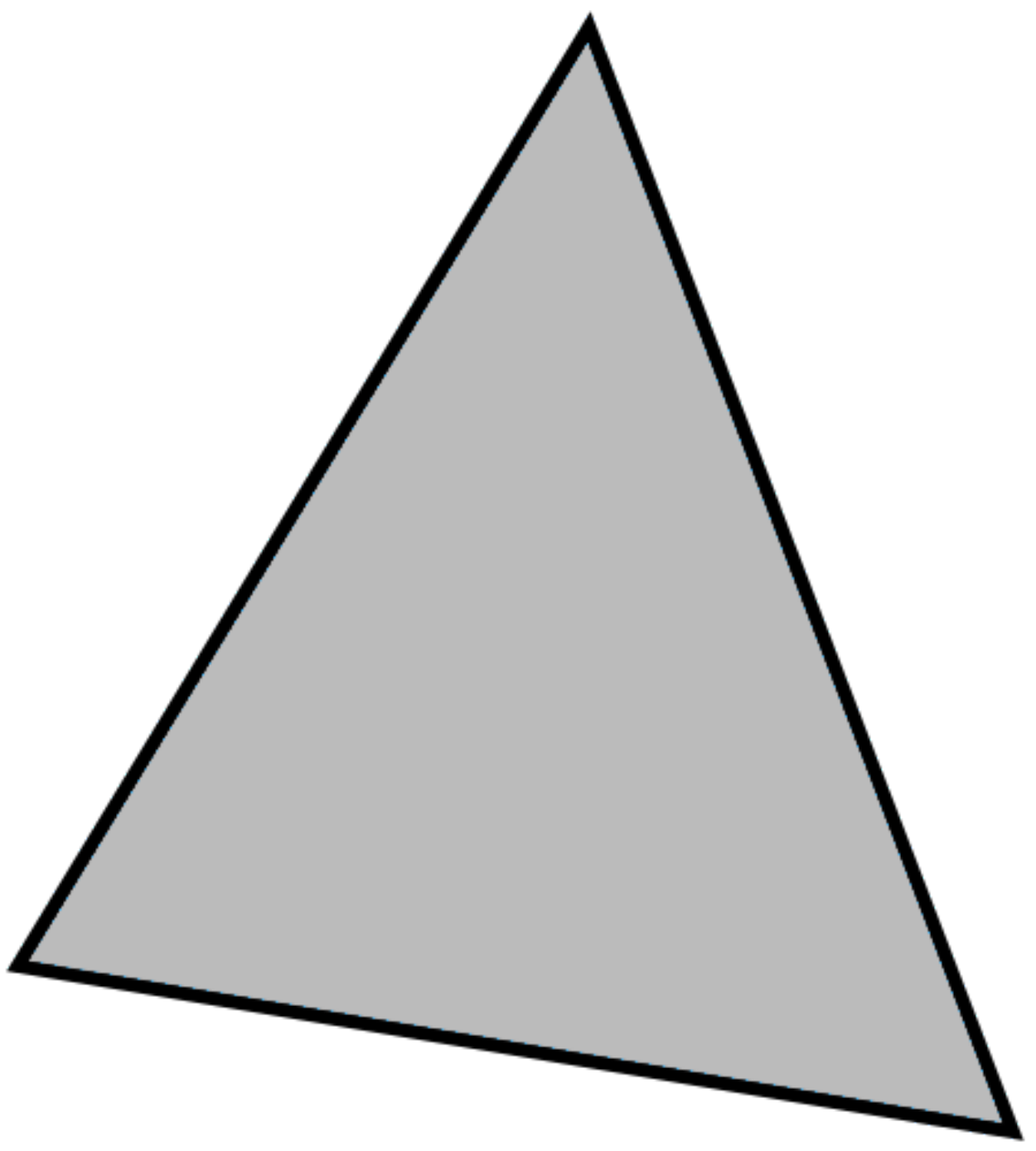}
\caption{Subgraphs that form a solid triangle cover complex}
\end{subfigure}
\begin{subfigure}{.96\linewidth}
\centering
\includegraphics[height=3cm]{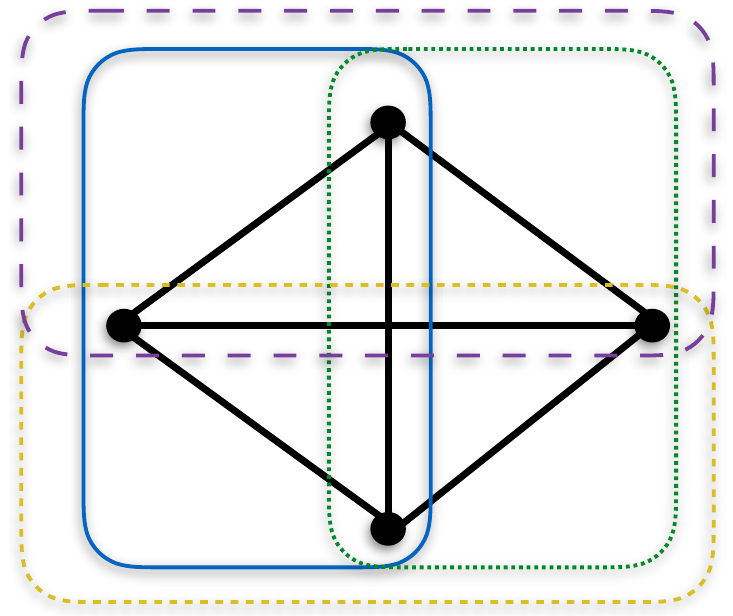}
\hspace{20pt}
\centering \includegraphics[height=3cm]{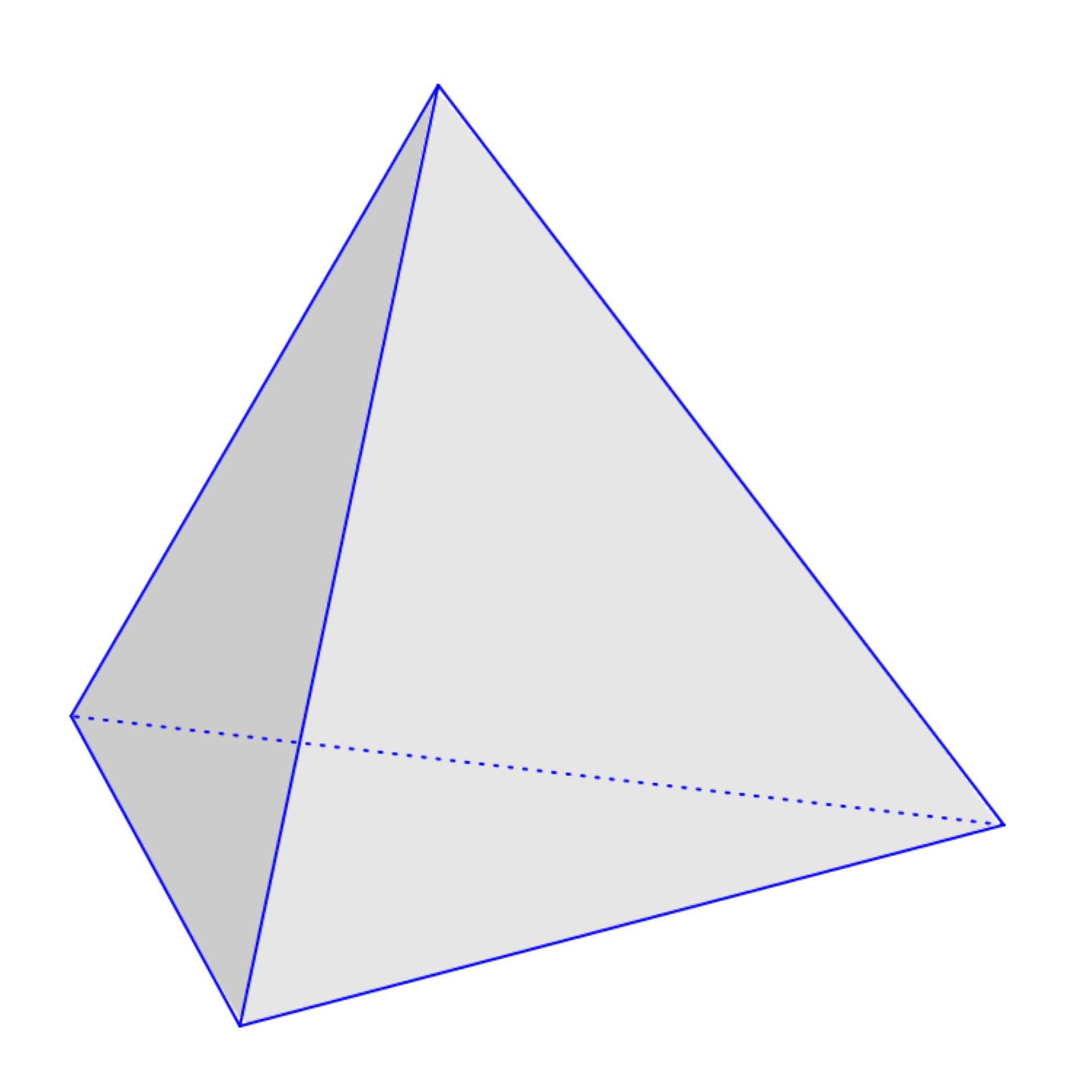}
\caption{Subgraphs that form an empty tetrahedron cover complex}
\end{subfigure}
\caption{Given local consistency, (a) is not globally consistent, while (b) and (c) are guaranteed to be globally consistent. (Left: the map graph with subgraphs circled out. Right: the corresponding cover complex.)}
\label{fig:cover_examples}
\end{figure}

Note that the $3^\tn{nd}$ condition in Theorem \ref{thm:consistency} is necessary. Figure~\ref{fig:cover_examples}(a) provides a simple counter example, which satisfies the $1^\tn{st}$ and $2^\tn{nd}$ conditions in Theorem \ref{thm:consistency}. The cover complex $\ck$, however, is homologous to the Torus $T^2$, which is not simply connected. It is easy to see that the local-consistency (which is trivial as each sub-graph is given by an edge) does not lead to the global consistency among these three edges.

Figure~\ref{fig:cover_examples}(c) provides another example to understand the correctness of Theorem~\ref{thm:consistency}. In this case, there are four objects. It is clear that enforcing the cycle-consistency among all four triple sub-graphs induces the cycle-consistency on the original graph. This argument aligns with Theorem~\ref{thm:consistency} as $\ck$ is simply connected. We defer detailed explanations to the supplementary material.

\begin{algorithm}[ht]
\caption{Greedy Construction of $\ck$}
\label{alg:greedyK}
\SetKwInOut{Input}{Input}
\SetKwInOut{Output}{Output}
\Input{Map graph $\cg = (\cv, \ce)$\\ Number of cover nodes $K$}
\Output{Cover nodes $\{\cg_i\}_1^K$}
Compute a graph clustering of $\cg$ to $K$ clusters (graph cut or $K$-means on graph embeddings)\\
Assign each of the cluster to $\cg_i,~i=1,\cdots,K$ \\
\While{condition not satisfied}{
%\tcc{inner-node update}
Assign $H_j$ to $\cg_i$ if $H_j$ is neighboring to $\cg_i$ in $\cg$ or within distance of $\epsilon$ to $\cg_i$ in the embedding space\\
Build $\ck$ from $\{\cg_i\}_i^K$ \\
Check if $\ck$ is connected \\
Compute homology group using \cite{Zomo2005} \\
Check if $\ch_1(\ck)$ is trivial \\
\If{Both conditions satisfied}{
Break
}
}
\end{algorithm}

%\subsection{Greedy Construction of $\ck$}
To develop an algorithm based on Theorem~\ref{thm:consistency}, we proposed a greedy algorithm to construct $\ck$ as in Algorithm \ref{alg:greedyK}. Note that a complex $\ck$ is \emph{simply connected}, if  1) it is connected; 2) the 1-dimensional homology group $\ch_1(\ck)$ is trivial. Condition 1) could be easily verified by any graph traversal algorithm (BFS/DFS), and condition 2) can be verified computationally as in~\cite{Zomo2005}.

\section{Distributed Optimization}
\label{sec:formulation}
%As has been shown in \cite{Huang2013, Chen2014, Zhou2015} the consistent graph matching problem could be solved nicely with a relaxed convex problem in a matrix completion manner. One drawback of their methods, however, is the difficulty to scale up as they formulate the problem to be solved in a global sense.

%Theorem \ref{thm:consistency} laid a good theoretical guarantee that we could instead of solving the global scale optimization, solving it in a distributed manner by iteratively solving the localized consistency problem of each $\cg_i$ and update globally by exchanging information of the joint consistency between $\cg_i$ and $\cg_j$. This will, as a consequence, break the problem of solving large scale global optimization into small pieces that could be done in a parallel and distributed way.
In this section, we introduce the proposed distributed formulation of recovering cycle-consistent maps from noisy pair-wise maps.

\subsection{Formulation}
%Given a map graph $\cg$ and a cover $\cv_\cg$, we formulate the distributed \emph{cycle consistency} problem from the observation in Theorem \ref{thm:consistency}. %Let $h(\cv_i)$ be the consistency objective function in node $\cv_i$. We would then need to find the optimal solution to minimize $\sum_{i} h(\cv_i)$ subject to the constraint that $\cv_i$ and $\cv_j$ are joint consistent.
Our formulation takes as input the pairwise base maps $\phi_{ij}$. We follow the state-of-the-art work on convex relaxation of enforcing cycle-consistent maps
\cite{Huang2013,Chen2014} to encode $\phi_{ij}$ into a data matrix $\bbx_{ij}$. Following the common strategy for optimizing point-based maps, we relax $\phi_{ij}$ to be a partial map and/or soft map, i.e. $\bbx_{ij}\in[0,1]^{m_i\times m_j}$, where $m_i$ denote the number of vertices in $H_i$. %where \tcb{$\bbx_{ij}(s,s')=1$} iff $(s,s')\in\phi_{ji}$.
%Here we relax the assumption that $\phi_{ij}$ is bijective, and in addition we allow partial maps. Hence $\bbx_{ij}$ is not necessarily square.

Let $\bbx_{\cv}$ be the matching matrix that encodes pair-wise maps in its blocks, i.e.
\[
\bbx_\cv =
\begin{pmatrix}
\mbi  & \bbx_{12} &  \cdots & \bbx_{1n} \\
\bbx_{21} & \mbi & \cdots & \vdots \\
\vdots & \cdots & \ddots & \vdots \\
\bbx_{n1} & \cdots & \cdots & \mbi \\
\end{pmatrix}
\]

%Following the objective in \cite{Zhou2015},
%Let $n$ be the number of \tcb{matching graphs}. %$m_i$ be the number of \tcb{vertices in matching graph} $H_i$. %\tcb{$\bbx_{ij}$ be the matrix for input base map $\phi_{ji}$, and $\bbx_\cv$ be the matrix of all input base maps}.
%$\bbx_\cv$ represents the noisy input in our formulation.
The goal of our formulation is to find a matrix $\mbx_\cv$ that encodes cycle-consistent maps from the noisy input $\bbx_\cv$.
%be the optimization variable which encodes the desired cycle consistency property.
To achieve this goal, one observation is that the desired matching matrix $\mbx_\cv$ is low-rank (c.f.~\cite{Huang2013}). Specififcally, we assume there is an universal object of size $m$, i.e. there are totally $m$ distinct entities for all the objects $H_i$'s in $\cv$. For each object $H_i$, we have a latent map encoded by $\mba_{H_i}\in \{0, 1\}^{m_i\times m}$, which maps a vertex from $H_i$ to an entity in the universal object. Let $\mba_\cv$ be a tall matrix that concatenates $\mba_{H_i}$, i.e., $\mba_\cv = (\mba_{H_1}^{T},\cdots, \mba_{H_n}^{T})^{T}$. It is easy to see that the map matrix $\mbx_\cv$ admits a low-rank factorization as $\mbx_\cv = \mb{A}_\cv \mb{A}_\cv^\top$. In~\cite{Huang2013, Zhou2015}, the authors use this property to develop robust algorithms for recovering $\mbx_\cv$ from noisy input maps.

Without losing generality. we assume $h(\mbx_\cv)$ is an objective function that measures the quality of a set of cycle-consistent maps encoded by $\mbx_\cv$, e.g., it promotes the low-rankness of $\mbx_\cv$. The precise expression of $h(\mbx_\cv)$ will be discussed later. Our distributed formulation is given by
\begin{equation}\label{eqn:dsdp}
\begin{array}{ll}
\min & \sum_i h\left(\mbx_{\cv_i}\right) \\
\st & \mbx_{\cv_{i\cap j}^i} = \mbx_{\cv_{i\cap j}^j}, \forall (i,j) \in \ce,
\end{array}
\end{equation}
where $\mbx_{\cv_{i\cap j}^i}$ is the matching matrix of $\cv_{i\cap j}$ in $\cv_i$, i.e. a sub-matrix of $\mbx_{\cv_i}$ by picking blocks that belong to the matching graphs in $\cv_{i\cap j}$. Each $h(\mbx_{\cv_i})$ indicates local consistency in $\cv_i$, and the condition that $\mbx_{\cv_{i\cap j}^i} = \mbx_{\cv_{i\cap j}^j}$ will guarantee that the overlapping subgraph are consistent. In such a manner, the consistency condition will propagate through the overlapping sub-graph to each component $\cv_i$ conceptually similar to our proof of Theorem \ref{thm:consistency}.

In the state-of-the-art methods of~\cite{Huang2013} and~\cite{Zhou2015}, the authors proposed different formulations of objective function $h(\mbx_\cv)$. We will use the formulation described in~\cite{Zhou2015} to demonstrate our framework, because of its competent performance and superior computational speed.

%We will leave in our supplementary material for the derivation using Huang's formulation in \cite{Huang2013}.

As in \cite{Zhou2015}, $h(\mbx_\cv)$ can be written as
\begin{equation}\label{eqn:gsdp}
\begin{array}{ll}
\min & \langle \mbw_\cv, \mbx_\cv\rangle + \lambda\|\mbx_\cv\|_* \\
\st & \mbx_\cv \succeq 0, \\
& \mbx_{\cv(ii)} = \mb{I}_{m_i}, \forall i\\
& \mbx_{\cv(ij)} = \mbx_{\cv(ji)}^\top, \forall i\neq j\\
& 0 \leq \mbx \leq 1\\
\end{array}
\end{equation}
where $\langle\cdot,\cdot\rangle$ is the matrix inner product, \tcb{$\|\cdot\|_*$ is the matrix nuclear norm}, and \tcb{$\mbw_\cv =  \alpha\mb{1} - \bbx_\cv$}, and $\mb{1}$ denote the matrix whose elements are $1$. The purpose of adding constant $\alpha$ is to impose a $L_1$ constraint on \tcb{$\mbx_\cv$} to promote sparsity.
%Note we abuse the notation slightly to overload the subscripts, namely to use $\mbx_{(ij)}$ to denote the $(i,j)^\tn{th}$ block of a matrix $\mbx$.
\tcb{We use $\mbx_{\cv(ij)}$ to denote the $(i,j)^\tn{th}$ block of the block matrix $\mbx_\cv$}.
As has been shown in \cite{Zhou2015}, the constraint $\mbx_\cv \succeq 0$ may be relaxed for a sufficiently large $\lambda$. Let $\cc_i$ encode the convex set induced by the constraints for $\cv_i$, we could then simplify the formulation of our distributed problem as
%of our optimization problem for $h(\cv_i)$ to be
%
%\[
%\begin{array}{ll}
%\min & \langle \mbw_{\cv_i}, \mbx_{\cv_i}\rangle + \lambda\|\mbx_{\cv_i}\|_* \\
%\st & \mbx_{\cv_i} \in \cc_i\\
%\end{array}
%\]
%
%Let $\mbx_{\cv_{i\cap j}^i}$ be the mapping matrix of $\cv_{i\cap j}$ in $\cv_i$, i.e. a submatrix of $\mbx_{\cv_i}$ by picking blocks that belongs to the matching graphs in $\cv_{i\cap j}$. From Theorem \ref{thm:consistency}, we could then formulate
\begin{equation}\label{eqn:dsdp2}
\begin{array}{ll}
\min &\tcb{\sum_i\left(\langle\mbw_{\cv_i}, \mbx_{\cv_i}\rangle + \lambda\|\mbx_{\cv_i}\|_*\right)} \\
\st & \mbx_{\cv_i} \in \cc_i\\
& \mbx_{\cv_{i\cap j}^i} = \mbx_{\cv_{i\cap j}^j}, \forall (i,j) \in \ce
\end{array}
\end{equation}

\section{Alternating minimization}
\label{sec:solver}
\subsection{Algorithms}
\label{ssect:alg}
The nuclear norm minimization in \eqref{eqn:dsdp2} can be efficiently optimized using recent results on low-rank matrix recovery, which directly enforce low-rank decompositions $\mbx_{\cv_i} = \mba_{\cv_i}\mbb_{\cv_i}^\top$ (c.f.~\cite{Cabral2013, Hastie2015, Zhou2015}). Here $\mba_{\cv_i}$ and $\mbb_{\cv_i}$ are latent variables. According to~\cite{Recht2010}, we can write the nuclear norm as
\[
\|\mbx\|_* = \min_{\mba, \mbb: \mba\mbb^\top = \mbx} \frac{1}{2}\left(\|\mba\|_F^2 + \|\mbb\|_F^2\right).
\]
To make the notations uncluttered, we will shorten $\mbx_{\cv_i}$ and $\mbx_{\cv_{i\cap j}}^i$ as $\mbx_i$ and $\mbx_{ij}$, respectively. %, the optimization problem in \eqref{eqn:dsdp} could be rewritten as
%\begin{equation}\label{eqn:dsdp-reform}
%\begin{array}{ll}
%\min & \sum_i\left(\langle \mbw_i, \mbx_i\rangle + \frac{\lambda}{2}\|\mba_i\|_F^2 + \frac{\lambda}{2}\|\mbb_i\|_F^2\right) \\
%\st & \mbx_i = \mba_i\mbb_i^\top\\
%& \mbx_{ij} = \mbx_{ji}, \forall (i,j) \in \ce
%\end{array}
%\end{equation}
%where we omit the constraints that $\mbx_i$ and $\mbx_{ij}$ has to %be in a convex set %$\cc_i$ and $\cc_{ij}$\footnote{\tcb{$\cc_{ij}$ %encodes the convex set for $\cv_i\cap\cv_j$ in a similar manner as %$\cc_i$ for $\cv_j$.}}
%respectively.
Moreover, let $\mbe_{ij}$ denote the selection matrix to extract the part of $\mbx_i$ that belongs to the set of $\cv_i\cap\cv_j$, i.e. %$\mbe_{ij} = \begin{bmatrix}\mb{e}_{i1} & \mb{e}_{i2} & \cdots & \mb{e}_{i|\cv_i\cap\cv_j|}\end{bmatrix}$, where
%\[
%\mb{e}_{ik} = \begin{bmatrix} 0 & \cdots & 0 & \underbrace{\mbi}_\text{$s$-th block} & 0 & \cdots & 0 \end{bmatrix}^\top.
%\]
%assuming we select $s$-th block out of $\mbx_i$ as the $k$-th block in $\mbx_{ij}$, namely
$\mbx_{ij} = \mbe_{ij}^\top\mbx_i\mbe_{ij}$. With this setup, the condition on intersection consistency %mapping within the intersections
%, $\mbx_{ij} = \mbx_{ji}$
becomes $\mbe_{ij}^\top\mbx_i\mbe_{ij} = \mbe_{ji}^\top\mbx_j\mbe_{ji}$.

We then finalize our formation of the problem in \eqref{eqn:dsdp} as
\begin{equation}\label{eqn:dsdp-final}
\begin{array}{ll}
\min & \sum_i\left(\langle \mbw_i, \mbx_i\rangle + \frac{\lambda}{2}\|\mba_i\|_F^2 + \frac{\lambda}{2}\|\mbb_i\|_F^2\right) \\
\st & \mbx_i = \mba_i\mbb_i^\top, \\
& \mbe_{ij}^\top\mbx_i\mbe_{ij} = \mbe_{ji}^\top\mbx_j\mbe_{ji}, \\
& \mbx_i \in \cc_i,
\end{array}
\end{equation}
In all our experiments, we set $\alpha=0.1$, $\lambda = 50$, $\mu = 64$, and $\beta=1$.

We apply ADMM to solve \eqref{eqn:dsdp-final}. The solver
%for \eqref{eqn:admm_lagrangian}
is summarized in Algorithm \ref{alg:dsdp}. In particular, $\mby_i$ and $\mbz_{ij}$ are dual variables. The constraints on $\mbx$ are handled implicitly and updated in the alternating algorithm. The ADMM algorithm updates primal variables by minimizing $\cl$ and then applies gradient descent to update the dual variables.
%using the augmented Lagrangian.
%is
%
%\begin{multline}
%\label{eqn:admm_lagrangian}
%\cl = \sum_i\left(\langle \mbw_i, \mbx_i\rangle + \frac{\lambda}{2}\|\mba_i\|_F^2 + \frac{\lambda}{2}\|\mbb_i\|_F^2 \right.\\
% + \left.\langle\mby_i, \mbx_i - \mba_i\mbb_i^\top\rangle + \frac{\mu}{2}\|\mbx_i - \mba_i\mbb_i^\top\|_F^2\right) \\
% + \sum_{i,j}\left(\langle \mbz_{ij}, \mbe_{ij}^\top\mbx_i\mbe_{ij} - \mbe_{ji}^\top\mbx_j\mbe_{ji}\rangle\right.\\
%  + \left.\frac{\beta}{2}\|\mbe_{ij}^\top\mbx_i\mbe_{ij} - \mbe_{ji}^\top\mbx_j\mbe_{ji}\|_F^2\right)
% %+ \sum_{i,j}\left( \langle\mbt_{ij}, \mbx_{ij} - \mbx_{ji}\rangle + \frac{\gamma}{2}\|\mbx_{ij} - \mbx_{ji}\|_F^2\right)
%\end{multline}
%
\begin{algorithm}[ht]
\caption{Distributed Graph Matching via ADMM}
\label{alg:dsdp}
\SetKwInOut{Input}{Input}
\SetKwInOut{Output}{Output}
\Input{Initial pairwise maps $\bbx_i$}
\Output{Consistent matches $\mbx_i$}
Initialize $\mba_i$, $\mbb_i$ randomly, and set $\mby_i$, $\mbz_{ij}$ to be $\mb{0}$ \\
$\mbw_i = \alpha\mb{1} - \bbx_i$ \\
\While{not converged}{
\tcc{inner-node update}
$\mba_i \leftarrow (\mbx_i + \frac{1}{\mu}\mby_i)\mbb_i(\mbb_i^\top\mbb_i + \frac{\lambda}{\mu}\mb{I})^\dagger$ \\
$\mbb_i \leftarrow (\mbx_i + \frac{1}{\mu}\mby_i)\mba_i(\mba_i^\top\mba_i + \frac{\lambda}{\mu}\mb{I})^\dagger$ \\
$\mbx_i \leftarrow \cp_{\cc_i}\left(\mbx_{i0}\right)$ \\
$\mby_i \leftarrow \mby_i^k + \mu\left(\mbx_i - \mba_i\mbb_i^\top\right)$ \\
$\mbz_{ij} \leftarrow \mbz_{ij}^k + \beta\left(\mbm_{i\rightarrow j}^{k+1} - \mbm_{j\rightarrow i}^{k+1}\right)$ \\
\tcc{inter-node information exchange}
node $j$ send $\mbm_{j\rightarrow i} = \mbe_{ji}^\top\mbx_j\mbe_{ji}$ to node i
}
Round $\mbx_i$ with a threshold of 0.5.
\end{algorithm}
Moreover, $\mba_i$ and $\mbb_i$ admit closed-form solution via solving least-squares. Moreover, $\mbx_{i0}$ is the solution to the linear equation
\begin{multline}
\mu\mbx_i + 2\beta\sum_j\mbe_{ij}\mbe_{ij}^\top\mbx_i\mbe_{ij}\mbe_{ij}^\top = \mu\mba_i\mbb_i^\top - (\mbw_i + \mby_i) \\
+ \sum_j\mbe_{ij}(2\beta\mbm_{j\rightarrow i}^k - \mbz_{ij} + \mbz_{ji})\mbe_{ij}^\top. \nonumber
\end{multline}
Furthermore, the update on $\mbx_i$ requires a projection onto the convex set $\cc$, $\cp_\cc(\cdot)$, i.e. $\cp_\cc(\mbx_0)$ is the solution to the problem
\[
\min_{\mbx\in\cc}\|\mbx - \mbx_0\|_F^2.
\]
This is essentially a linear programming problem, and can be solved efficiently. We refer to the supplementary material for a derivation.

The key point in Algorithm \ref{alg:dsdp} is the separation of inner-node update and inter-node information exchange. It can be seen that all the matrix computations are done in each node separately, which indicates a distributed computation. While after each iteration, adjacent nodes will need to exchange information by passing messages. Namely, for node $\cv_j$ and all its neighboring nodes $\cv_i$, a message will be sent in the form of $\mbm_{j\rightarrow i} = \mbe_{ji}^\top\mbx_j\mbe_{ji}$. Note that there is no overhead on generating these messages. Recall that $\mbe_{ji}$ is just a sub-block extraction matrix, and the way $\mbm_{j\rightarrow i}$ is computed is simply by extracting the sub-block of the matching matrix $\mbx_j$ that belongs to the intersection $\cv_i\cap\cv_j$. Since all the computations are indeed done on each node individually, the proposed algorithm is essentially completely distributed, and the only add-on is a syncing stage.

%\subsection{Global Recovery}
%Algorithm \ref{alg:dsdp} only recovered partially $\mbx$ in the global sense, i.e. it only recovers consistent matchings between pairs of graphs that are within the same cover node, namely $\mbx_i$'s. It, however, comes with a factorization of $\mbx_i$'s readily available, $\mbx_i=\mba_i\mbb_i^\top$. Therefore, if the global recovery is the ultimate goal, it could be further done with this global formulation
%\begin{equation}\label{eqn:global}
%\begin{array}{ll}
%\min & \|\mba_i-\mbe_i^\top\mba\mbe_i\|_F^2 + \|\mbb_i-\mbe_i^\top\mbb\mbe_i\|_F^2 + \gamma\|\mbx\|_*\\
%\st & \mbx = \mba\mbb^\top\\
%& \mbx \in \cc
%\end{array}
%\end{equation}
%where $\mbe_i$ is the selection matrix similar to $\mbe_{ij}$ in Section \ref{ssect:alg}.
%
%This problem turns out to be separable in terms of $\mba$ and $\mbb$, where each comes with a least square solution. The global matching matrix $\mbx$ could then be recovered by $\mbx = \cp_\cc(\mba\mbb^\top)$ with a thresholding.

\subsection{Complexity}
The computational complexity of Algorithm \ref{alg:dsdp} is dominated by matrix multiplication. In our approach, the per-iteration complexity is controlled by the leading node in $\cg$, i.e. $O\left(\max_i(\sum_{H_j\in\cv_i}m_j)^2\max_i(|\cv_i|)\right)$, where $|\cv_i|$ is the total number of distinct entities among all the objects in $\cv_i$. In contrast, the per-iteration complexity of~\cite{Chen2014} and~\cite{Zhou2015} is $O((\sum_im_i)^3)$ and $O((\sum_im_i)^2m)$ respectively. Furthermore, in our experiments, we found the total number of iterations to converge for our method is comparable to that of ~\cite{Zhou2015}.

%It is clear that our proposed method is well parallelizable, such that all the computations are done in distributed computing unit, and the only information exchange are the messages $\mbm_{j\rightarrow i}$ and $\mbm_{i\rightarrow j}$ passed between adjacent covers $\cv_i$ and $\cv_j$.

\section{Experiments}
\label{sec:exp}
\subsection{Simulation}

In this section, we perform experimental evaluation using synthetic datasets.

We followed the same experimental setup as in \cite{Chen2014, Zhou2015}. Given an optimized matching matrix $\mbx^*$ and the ground truth mapping $\mbx_g$, we access the quality of $\mbx^*$ by measuring the intersection over union (or IOU) score:
\[
1 - \frac{|\tau(\mbx^*)\cap\tau(\mbx_g)|}{|\tau(\mbx^*)\cup\tau(\mbx_g)|}
\]
where $\tau(\cdot)$ denotes the mapping induced from the matching matrix, and $|\cdot|$ denotes the set size. Note in our distributed setting, we could only partially recover $\mbx^*$ given $\mbx_i^*$. Therefore, our ground truth setting is also different from \cite{Chen2014, Zhou2015} in this regard.

\subsubsection{Matching Errors}
\label{sssect:merror}

In the first experiment, we aimed to evaluate the matching performance between our algorithm, DMatch, and the global algorithm, MatchALS, as in \cite{Zhou2015}. We considered the following model to generate the testing examples. The total number of graphs is denoted by $n$. The size of the universe is fixed at $r=20$ points. In each graph, a point is randomly observed with a probability $\rho_0$. We simulated error corruption by randomly removing true mapping and adding false ones with a corruption rate $\rho_e$.

We considered two settings in our experiments. In the first setting, we constructed our cover graph by making a sparse three way tree. This was done by randomly selecting a subset of $\cv$ as a common intersection $\cv_c$ and then split the rest evenly into the three cover nodes $\cv'_1,\cv'_2,\cv'_3$. As a consequence, each cover node is equal to $\cv_i = \cv_c\cup\cv'_i$. %In addition, we compare our results when the overlaps between cover nodes are dense, namely, we increase the size of the overlap between cover nodes.
In the second setting, we increased the overlap density by circularly adding one more split to each cover, i.e. $\cv_i = \cv_c\cup\cv'_i\cup\cv'_{i+1}$. We compared DMatch to MatchALS by varying the parameters $\rho_0$, $\rho_e$, and $n$. For both algorithms, we used $m=2r$ and $\lambda=50$.

\begin{figure}[ht]
\centering
\begin{subfigure}{.96\linewidth}
\begin{subfigure}{.32\linewidth} \centering \includegraphics[width=0.96\linewidth]{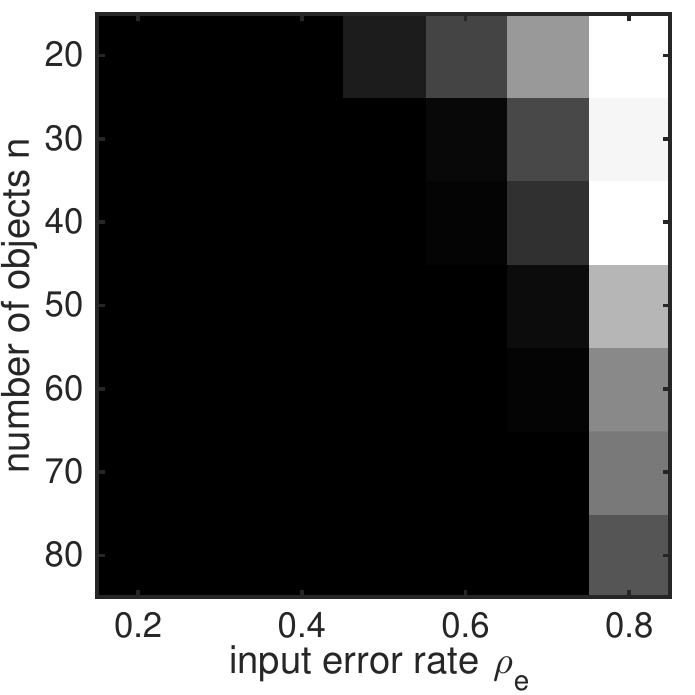}
\vspace{-10pt}\center{\footnotesize DMatch-sparse}
\end{subfigure}
\begin{subfigure}{.32\linewidth} \centering \includegraphics[width=0.96\linewidth]{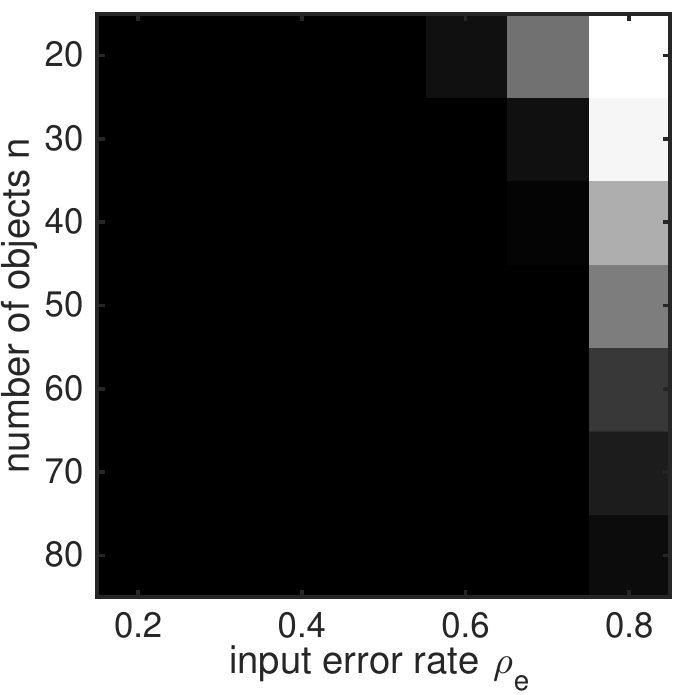}
\vspace{-10pt}\center{\footnotesize DMatch-dense}
\end{subfigure}
\begin{subfigure}{.32\linewidth} \centering \includegraphics[width=0.96\linewidth]{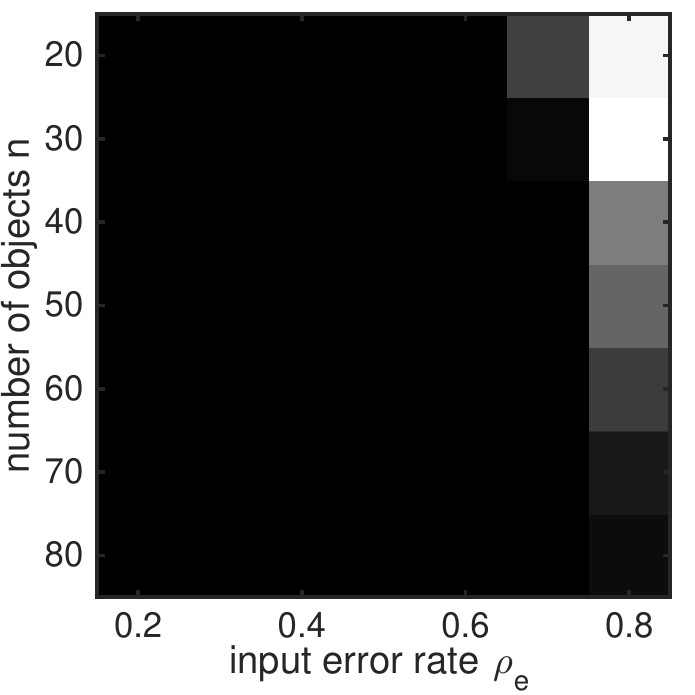}%{sparse_match_error_obs}
\vspace{-10pt}\center{\footnotesize MatchALS}
\end{subfigure}
\caption{number of objects $n$ + input error rate $\rho_e$}
\end{subfigure}
\centering
\begin{subfigure}{.96\linewidth}
\begin{subfigure}{.32\linewidth} \centering \includegraphics[width=0.96\linewidth]{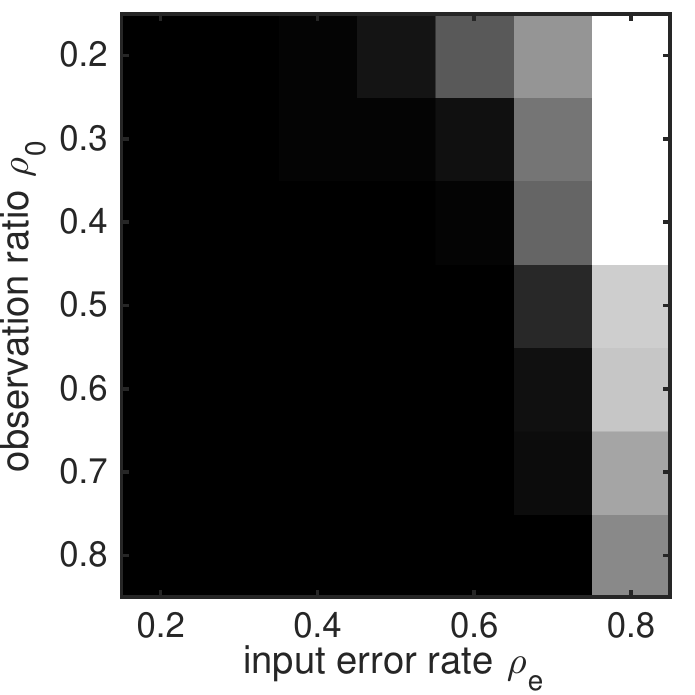}
\vspace{-10pt}\center{\footnotesize DMatch-sparse}
\end{subfigure}
\begin{subfigure}{.32\linewidth} \centering \includegraphics[width=0.96\linewidth]{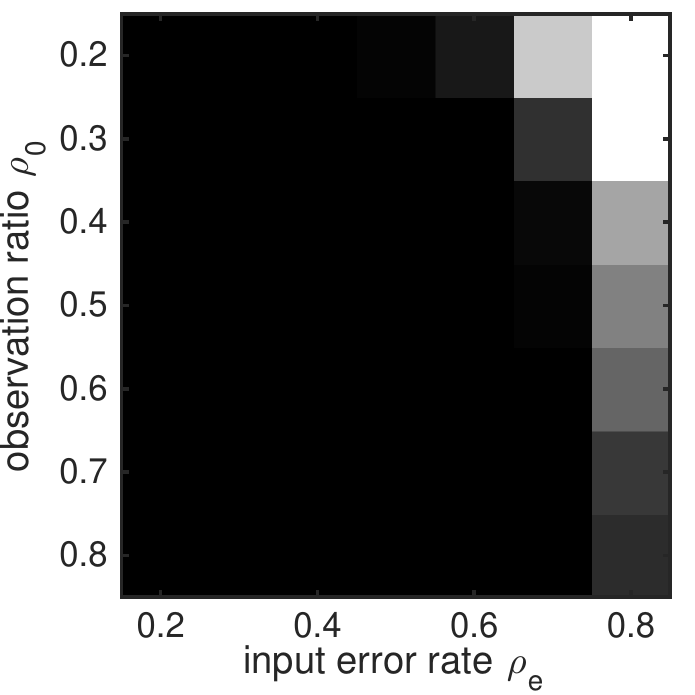}
\vspace{-10pt}\center{\footnotesize DMatch-dense}
\end{subfigure}
\begin{subfigure}
{.32\linewidth} \centering \includegraphics[width=0.96\linewidth]{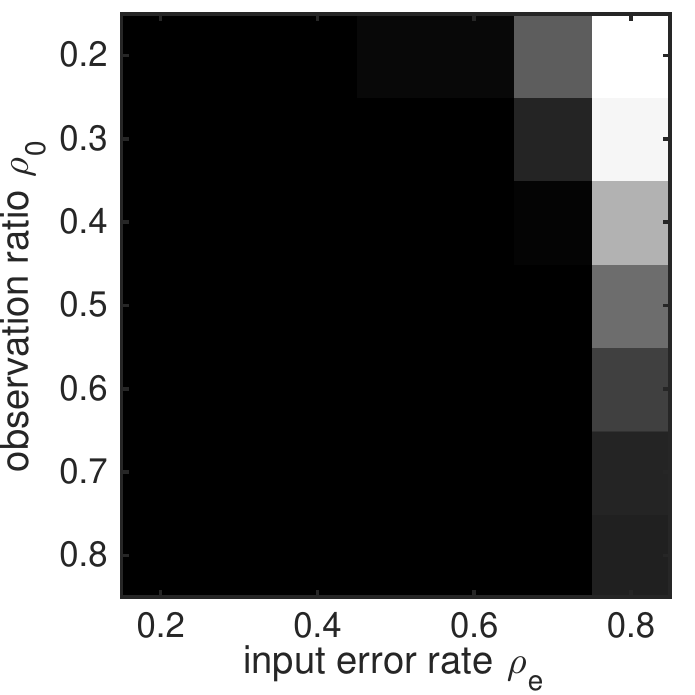}
\vspace{-10pt}\center{\footnotesize MatchALS}
\end{subfigure}
\caption{observation ratio $\rho_0$ + input error rate $\rho_e$}
\end{subfigure}
%\centering
%\begin{subfigure}{.96\linewidth}
%\begin{subfigure}{.48\linewidth} \centering \includegraphics[width=0.96\linewidth]{m_match_error_n}
%\end{subfigure}
%\begin{subfigure}{.48\linewidth} \centering \includegraphics[width=0.96\linewidth]{m_match_error_obs}
%\end{subfigure}
%\caption{MatchALS}
%\end{subfigure}
\caption{Matching error comparison. Darker color means lower matching error. The fixed parameter is set to $\rho_0=0.6$ and $n = 50$ respectively.}
\label{fig:merror}
\end{figure}

Figure~\ref{fig:merror} shows matching errors under various configurations, for both DMatch and MatchALS. In general, lowering input error and increasing observation ratio or increasing the total number of objects will improve the matching performance, i.e. with a lower matching error. In addition, we can see that increasing the overlap between cover nodes would have a positive impact on the recovery (comparison between Figure \ref{fig:merror}(a) and Figure \ref{fig:merror}(b)). %This results is inevitably desired as increasing the size of the overlap will help include more information between pairs of objects, and hence will produce better results.

Furthermore, in a comparison between Figure \ref{fig:merror}(b) and Figure \ref{fig:merror}(c), we can see that when the cover is dense enough, i.e. the size of the overlaps are sufficiently large, the matching error would approach that of MatchALS, which is the global recovery.

\subsubsection{Graph Covers}
In the second experiment, we aimed to understand more on the effects of graph covers. Specifically, we construct a ground truth graph cover by selecting a sparse cover as in Section \ref{sssect:merror}. For every pair of graphs within the same cover node, we set the error rate to be $\rho_{in}$, and for every pair between different cover node, we set the error rate to be $\rho_{out}$. The experiment is then conducted by comparing DMatch to 1) using the ground truth cover and 2) using a randomly constructed cover.

\begin{figure}[ht]
\centering
\begin{subfigure}{.49\linewidth}
\centering
\begin{subfigure}{.49\linewidth} \centering \includegraphics[width=0.96\linewidth]{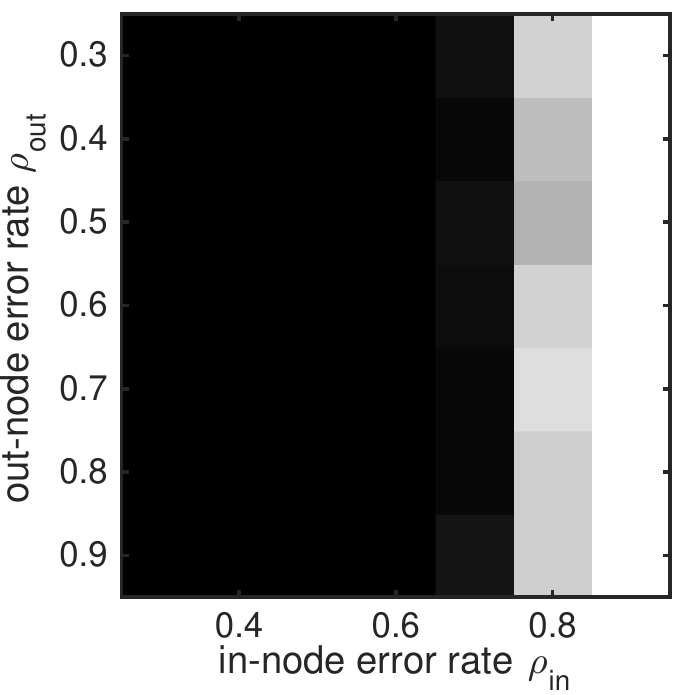}
\centering{\footnotesize sparse}
\end{subfigure}
\begin{subfigure}{.49\linewidth} \centering \includegraphics[width=0.96\linewidth]{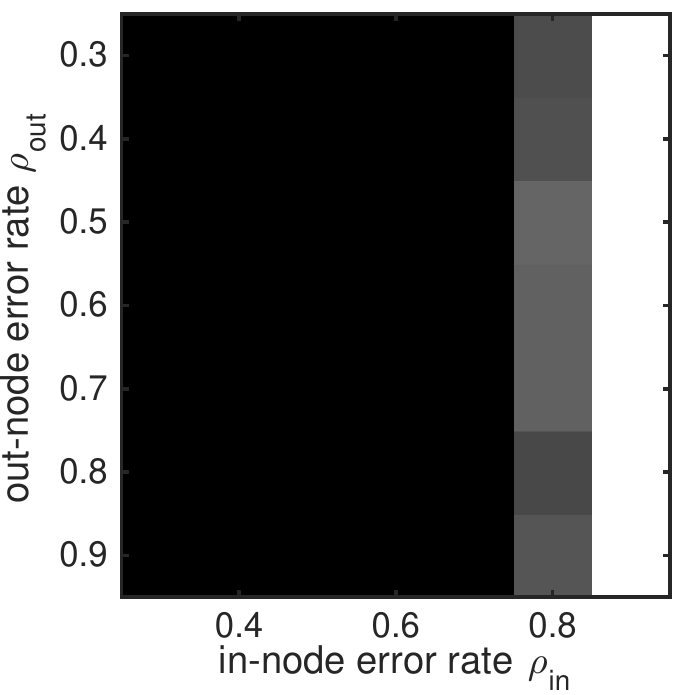}
\centering{\footnotesize dense}
\end{subfigure}
\caption{\footnotesize ground truth cover}
\end{subfigure}
\centering
\begin{subfigure}{.49\linewidth}
\centering
\begin{subfigure}{.49\linewidth} \centering \includegraphics[width=0.96\linewidth]{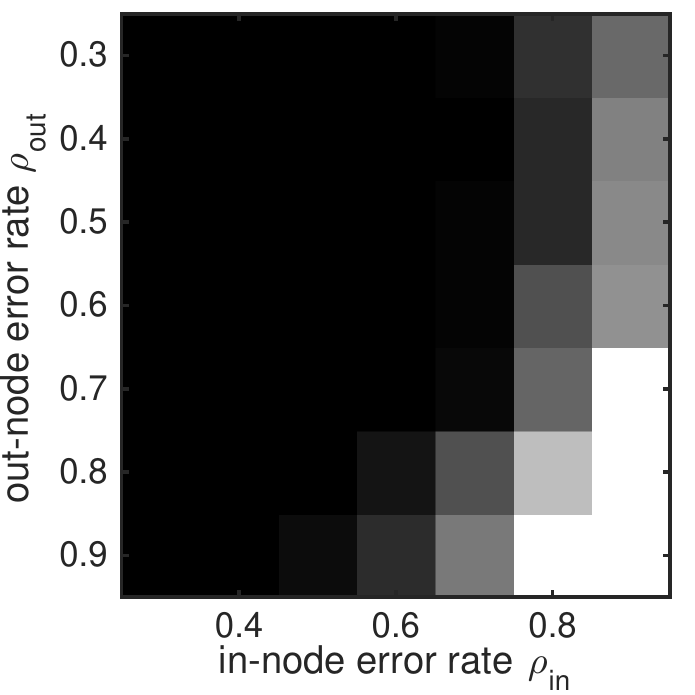}
\centering{\footnotesize sparse}
\end{subfigure}
\begin{subfigure}{.49\linewidth} \centering \includegraphics[width=0.96\linewidth]{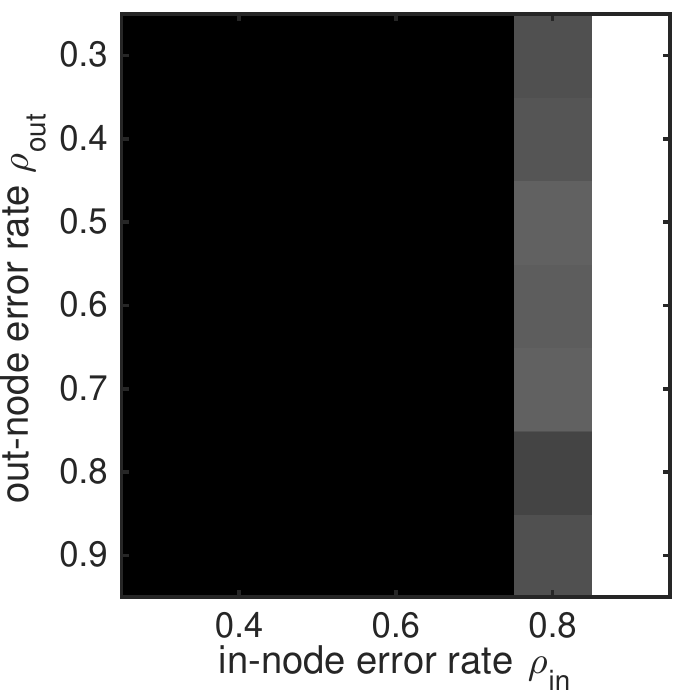}
\centering{\footnotesize dense}
\end{subfigure}
\caption{\footnotesize randomly constructed cover}
\end{subfigure}
\caption{Effects on recovery from the construction of graph cover. Darker color means lower matching error. The fixed parameter is set to $\rho_0=0.6$ and $n = 50$.}
\label{fig:gcover_error}
\end{figure}

Figure~\ref{fig:gcover_error} shows the experimental results. In Figure~\ref{fig:gcover_error}(a), we can see that using the ground truth cover, the matching performance does not depend much on $\rho_{out}$, because we explicitly disregarded any information from $\rho_{out}$. While on the randomly constructed cover, the two error rates are mixed up. Specifically, the results favor more on small $\rho_{in}$ and small $\rho_{out}$ at the same time. This situation, however, starts to change when the cover becomes denser, i.e. the increasing overlap between cover nodes. The dependency on $\rho_{out}$ disappears as shown in Figure \ref{fig:gcover_error}(b). One potential explanation is that with the increasing overlap between cover nodes, the portion of out-node pairs become smaller, as well as the resultant portion of induced error from them. As a consequence, the mixed error rate is dominated by $\rho_{in}$. In addition, we could also see that with denser cover, the algorithm is more error tolerant. This comes with a trade-off that on average the size of each cover becomes bigger and the computational cost also increases.

\subsection{Real Experiments}

\subsubsection{CMU House Sequence}
In this part of the experiment, we intend to test the scalability of our distributed algorithm. We use the CMU House sequence\footnote{\url{http://vasc.ri.cmu.edu//idb/html/motion/house/index.html}} as the testing images. This sequence has been widely used to test different graph matching algorithms. It consists of 110 frames, and there are 30 feature points labeled consistently across all frames. We used the geometry based constraint in pairwise matching as is done in \cite{Hu2014}. To construct a valid cover complex $\ck$, we first computed all pairwise matches and for each match, the result is encoded using a binary matrix. We then built a fully connected matching quality graph, where each vertex represents an image, and the edge weight represents the matching score associated with each image pair. Since we knew the sequence was roughly generated by moving a camera in a single dimension, we computed the Laplacian embedding of the vertices onto 1 dimensional space using Fielder vector, and then we applied Algorithm \ref{alg:greedyK} to build the cover complex.
%run $K$-mean clustering on top of the Fielder vector. To compute an overlapping cover, we assign each vertex to all clusters that are within 1.5 times the distance to its assigned cluster center in the Fielder vector.
%Because the Fielder vector of the graph describes the connectivity of it, we split the range of the Fielder vector into 5 overlapping intervals where we try to balance the number of images contained within each interval. In this way, we build a line-structured cover complex with 5 cover nodes, and the max number of images contained in all the covers are 45.
We compared our DMatch algorithm with the global methods, MatchALS~\cite{Zhou2015}, MatchLift~\cite{Chen2014}, Spectral ~\cite{Pachauri2013}, and the distributed method described in~\cite{Leonardox2017}. Besides running global algorithms on fully connected map graph, we also ran them on the sparse map graph induced by our cover complex, marked with different $K$ values. The algorithm was implemented in a single laptop with Intel Core i7 2.0GHz CPU and 16GB DDR3 Memory. We measured the time used in each cover node separately and then took the max as the total computational time per iteration, where we assumed the cost for messages passed between adjacent cover nodes was negligible.

\begin{table}[ht]
\centering
  \begin{tabular}{ r r r r}
    \hline \hline
      & Error Rate & Iterations & Time \\ \hline
    Original & 0.1445 & - & - \\
    MatchALS & 0.1031 & 266 & 98.8 \\
    MatchLift & 0.1027 & 1000 & 3791.1 \\
    Spectral & 0.1277 & - & \bf{0.6} \\
    MatchALS ($K=4$) & 0.0161 & 380 &103.5 \\
    MatchALS ($K=6$) & 0.0648 & 1000 & 268.2 \\
    MatchLift ($K=4$) & \bf{0} & 1000 &4066.1 \\
    MatchLift ($K=6$) & \bf{0} & 1000 & 3972.3 \\
    DMatch ($K=4$) & \bf{0} & 203 & 28.9 \\
    DMatch ($K=6$) & \bf{0} & \bf{150} & 7.6 \\
    Consensus ($K=4$) & \bf{0} & 150 & 28.3 \\
    Consensus ($K=6$) & 0.0071 & 150 & 18.7 \\
    \hline \hline
  \end{tabular}
\caption{The error rate and the total computational time (seconds) on CMU House sequence.}
\label{tab:scalability}
\end{table}

Table~\ref{tab:scalability} shows the matching accuracy, timing and iterations used in these algorithms. In our experiments, Spectral method appeared to be the fastest method among all the methods. On the other hand, it has the largest reported error. The consensus algorithm is not error-driven, hence we used a preset 150 iterations to match the lowest number of iterations among all experiments. In general, distributed algorithms used less iterations to converge, and achieved at least an order of magnitude speed-up compared with global methods, while maintaining an error rate of {\bf 0} (almost 0 for Consensus algorithm with $K=6$). It can also be seen that for distributed algorithms once we increased the number of clusters from $K=4$ to $K=6$, both the number of iterations and total computational time decreased. This, on the other hand, proved our complexity analysis, since increasing the number of clusters would in general reduce the number of vertices in each cluster. Although both DMatch and Consensus algorithms achieved similar results in terms of accuracy and time, the latter requires knowledge of the number of universal entities and has limitations dealing with partial matches. Another interesting observation is after convergence, the error of MatchALS on the sparse map graph induced by our cover complex is reduced, when compared with that on the fully connected map graph. One explanation is that the graph cover structure grouped together images that have high pairwise matching quality and explicitly disregard any pairwise matches that are of low quality (covers are joint normal), and as a consequence, it is robust against noisy pair-wise matches. In addition, we also extended the experiments for $K=8,10$ and found the running time was not reduced significantly, in comparison with that from $K=4$ to $K=6$. Since we need to have a reasonable amount of overlaps between clusters to pass matching information around, increasing the number of clusters does not necessarily reduce the maximum size of the clusters.

\begin{figure*}[t]
\begin{subfigure}{.32\linewidth} \centering \includegraphics[width=0.96\linewidth]
%{graffiti_curve_composite}
{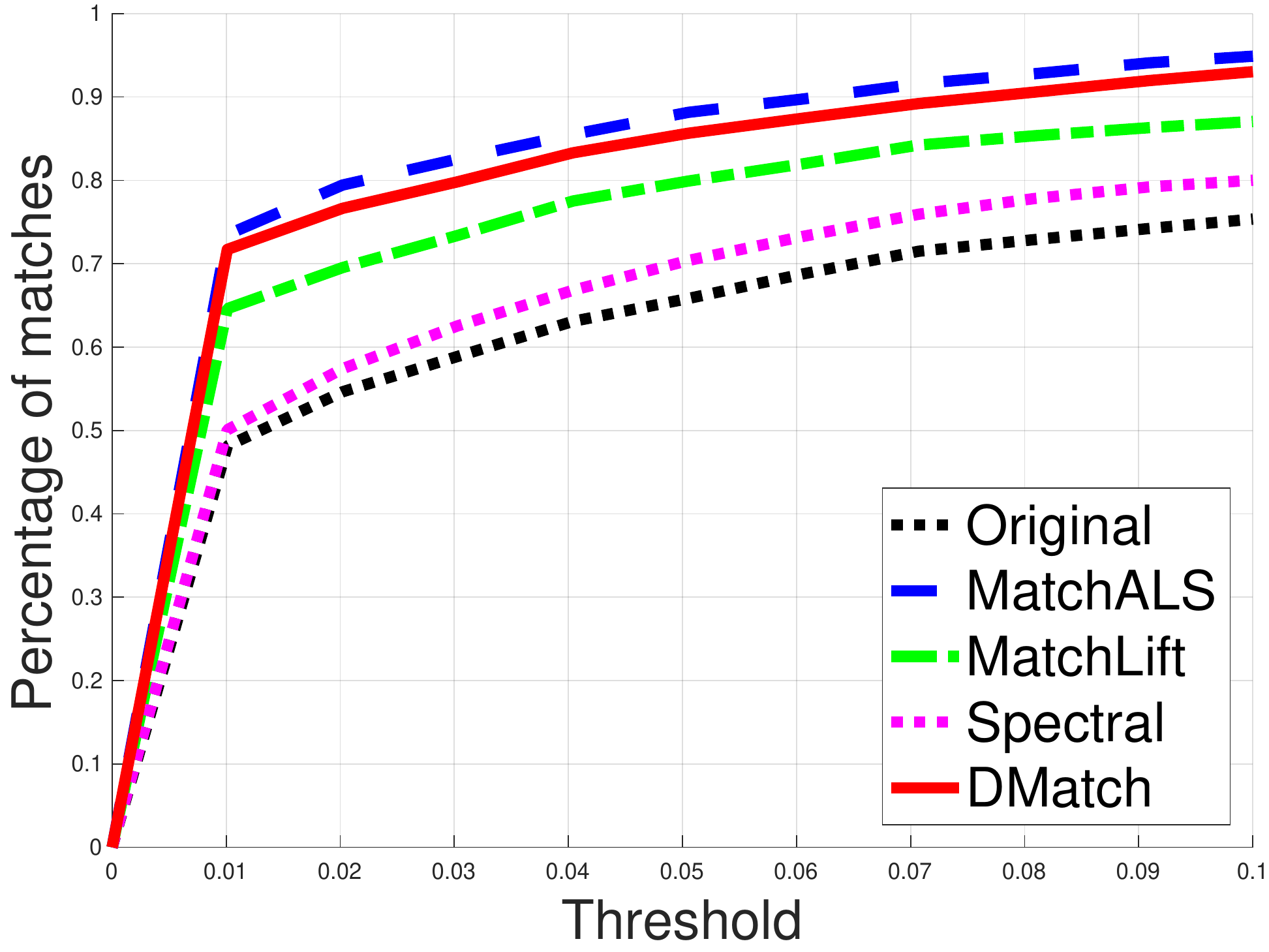}
\caption{Graffiti}
\end{subfigure}
\begin{subfigure}{.32\linewidth} \centering \includegraphics[width=0.96\linewidth]
%{bikes_curve_composite}
{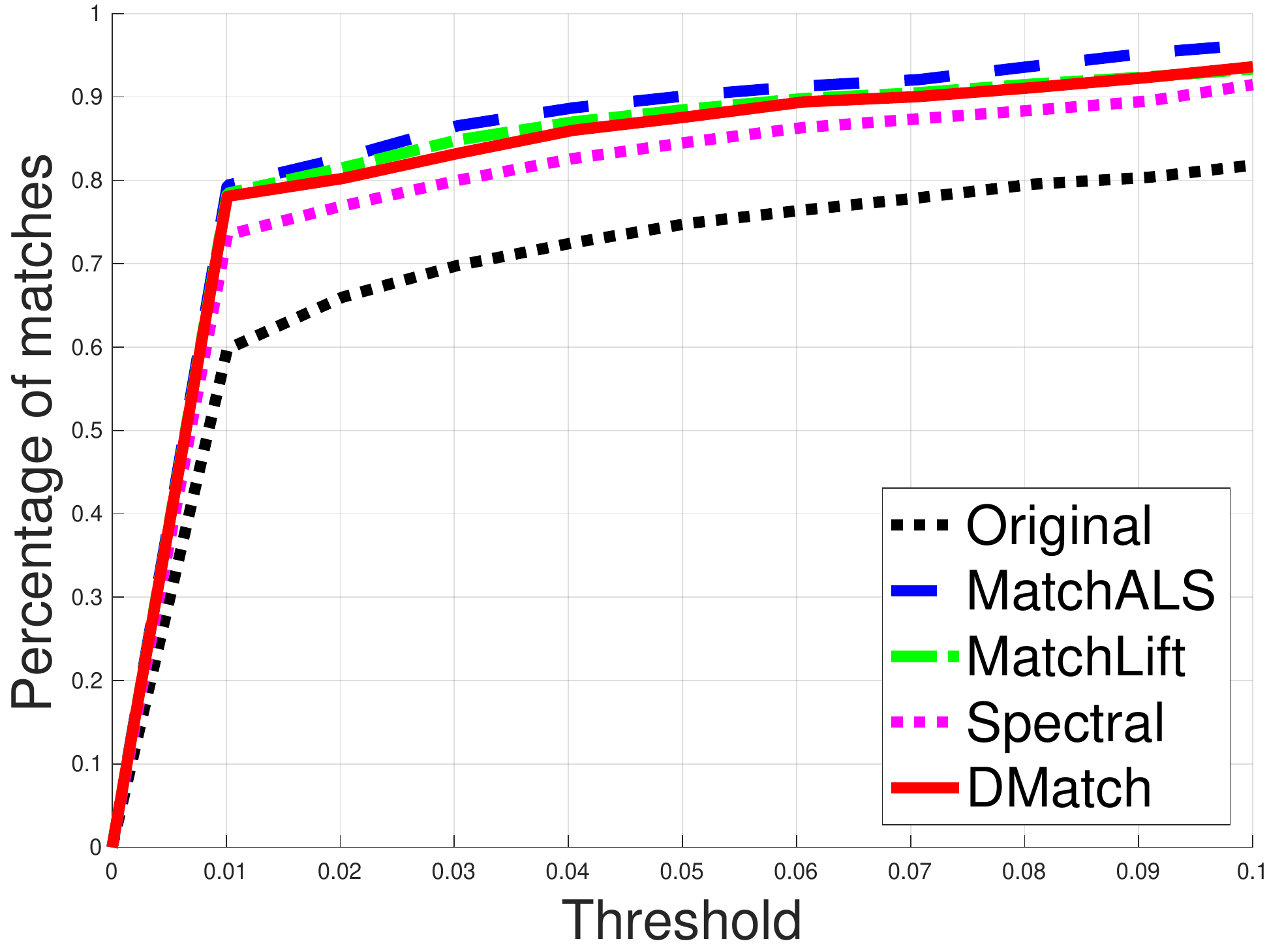}
\caption{Bikes}
\end{subfigure}
\begin{subfigure}{.32\linewidth} \centering \includegraphics[width=0.96\linewidth]
%{light_curve_composite}
{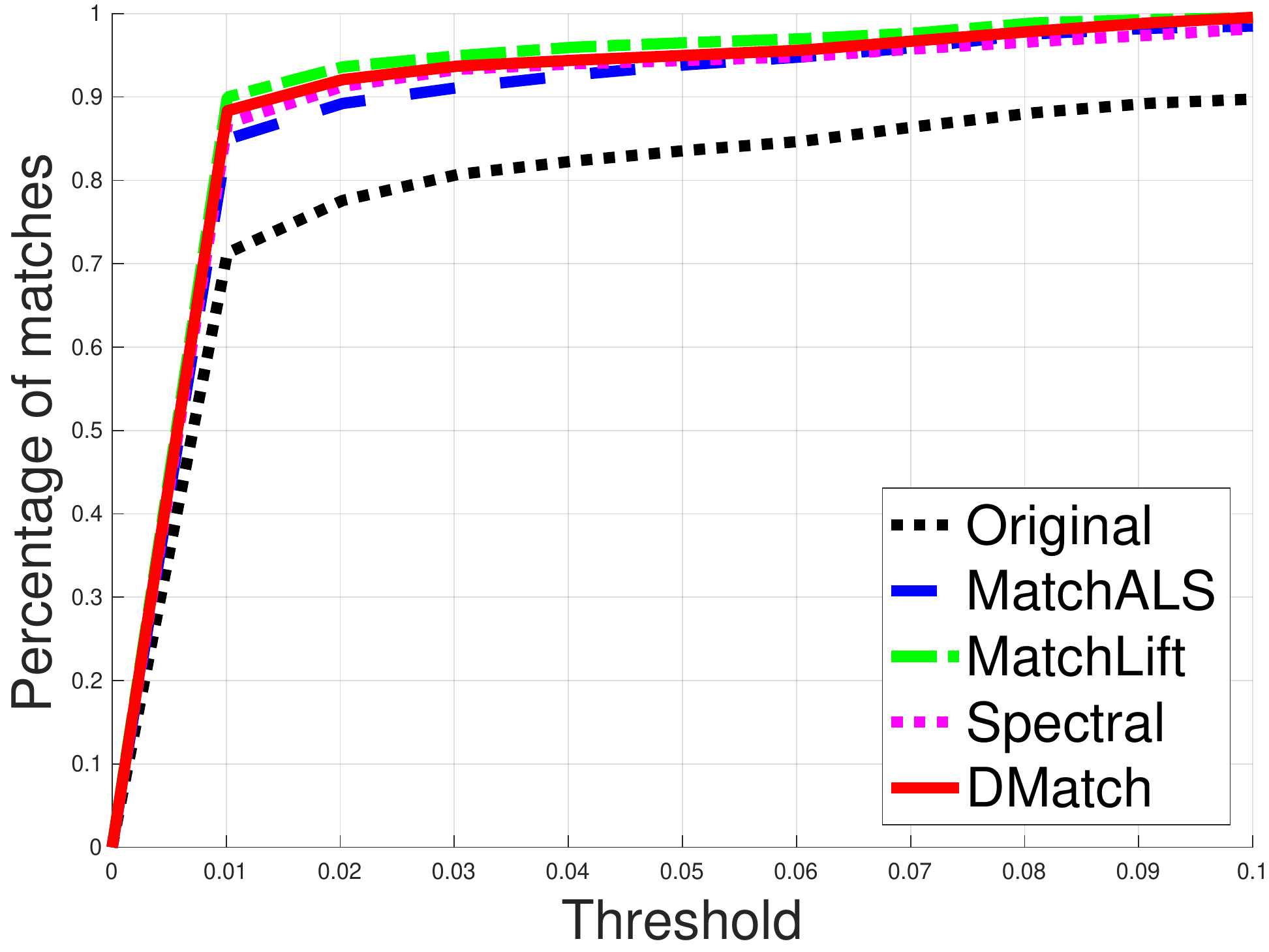}
\caption{Leuven}
\end{subfigure}
\caption{The performance curve on Graffiti, Bikes and Leuven datasets. The $y$-axis is the correct match ration and the $x$-axis is the threshold value over the image width. We compare DMatch (red solid) with MatchALS \cite{Zhou2015} (blue dashed), MatchLift \cite{Huang2013}, Spectral \cite{Pachauri2013}, and original pairwise matching (black dotted).}
\label{fig:error_curve}
\end{figure*}

\begin{figure*}[t]
\begin{subfigure}{.3\linewidth} \centering \includegraphics[height=6cm]{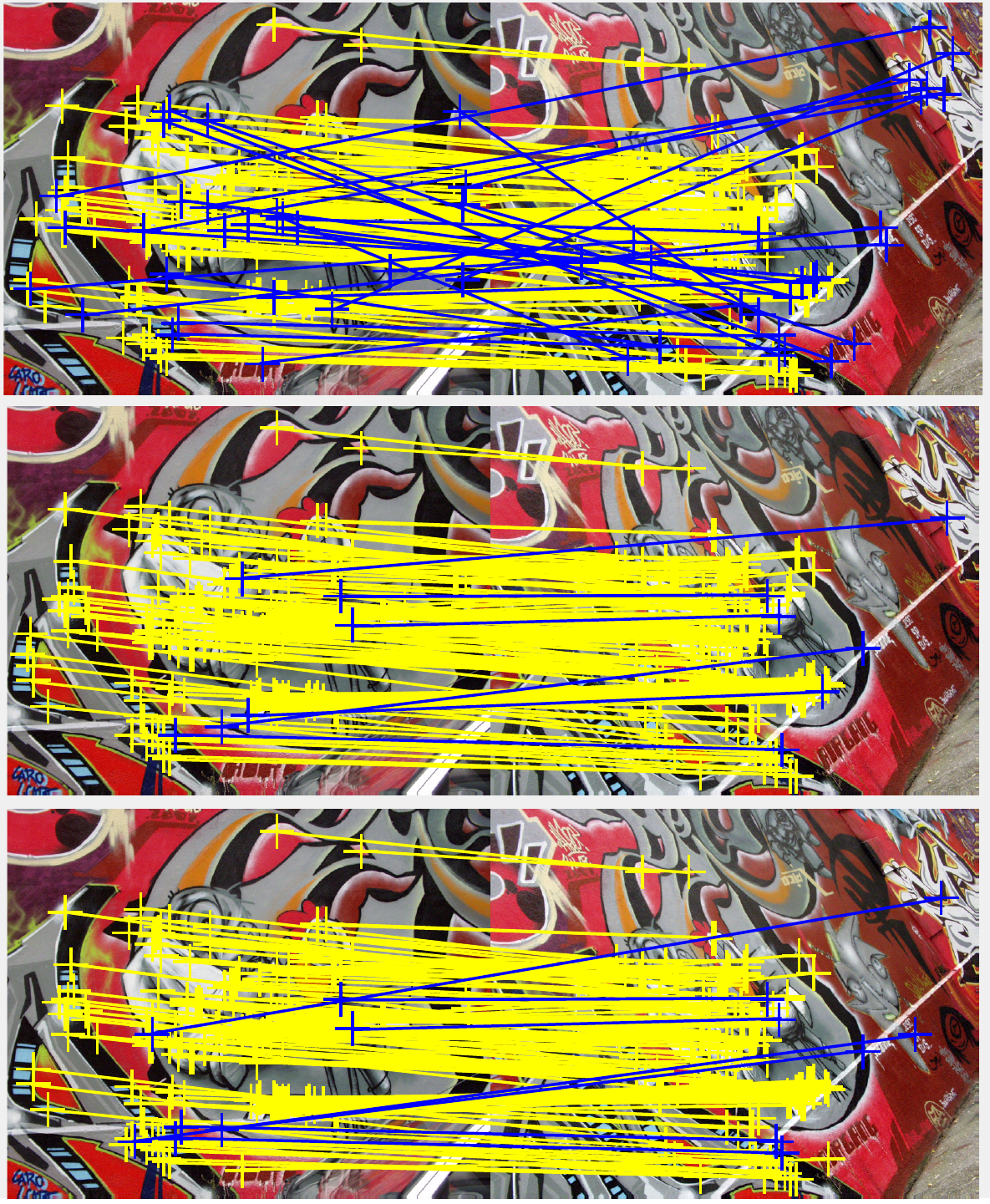}
\caption{Graffiti}
\end{subfigure}
\begin{subfigure}{.34\linewidth} \centering \includegraphics[height=6cm]{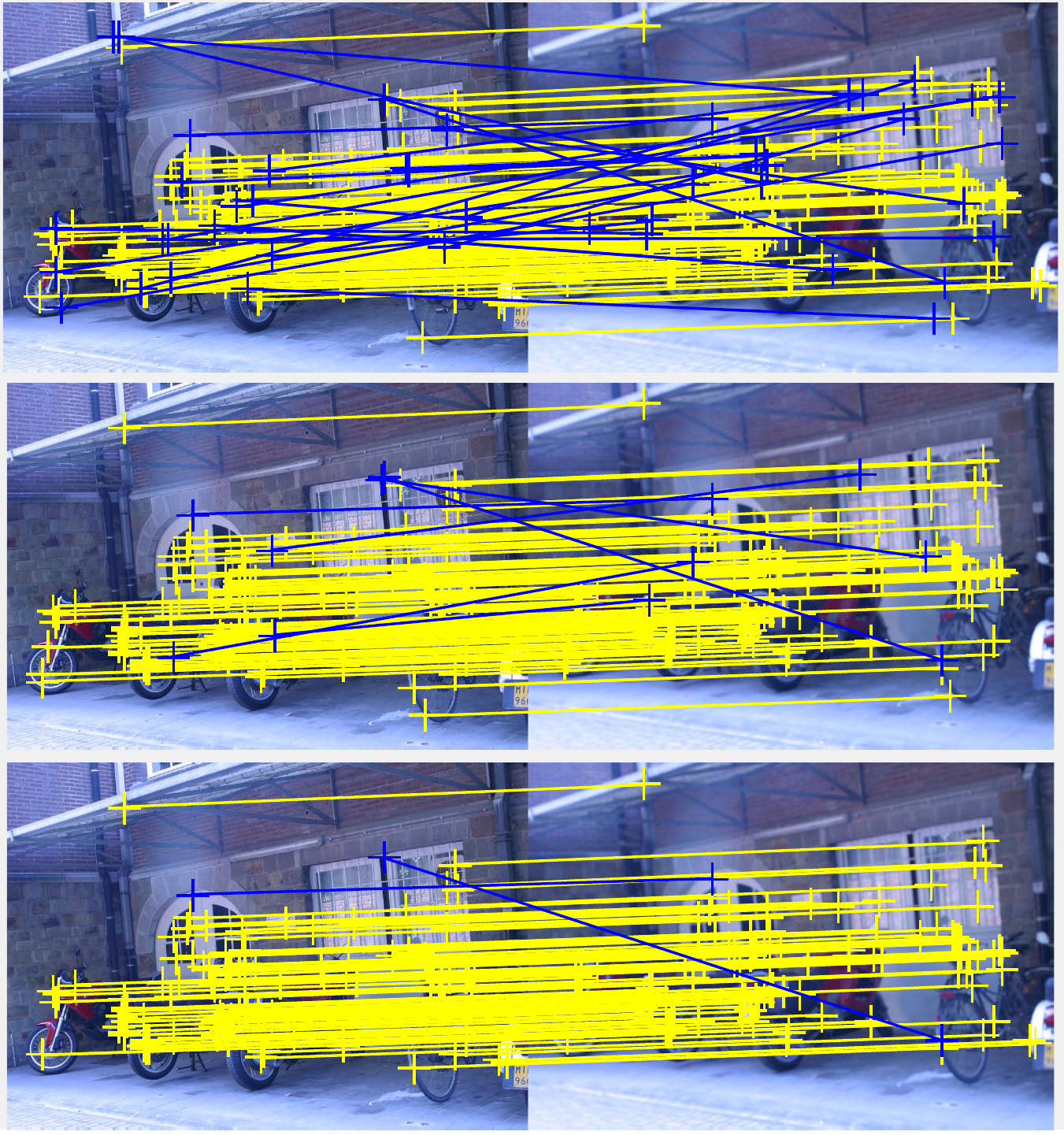}
\caption{Bikes}
\end{subfigure}
\begin{subfigure}{.34\linewidth} \centering \includegraphics[height=6cm]{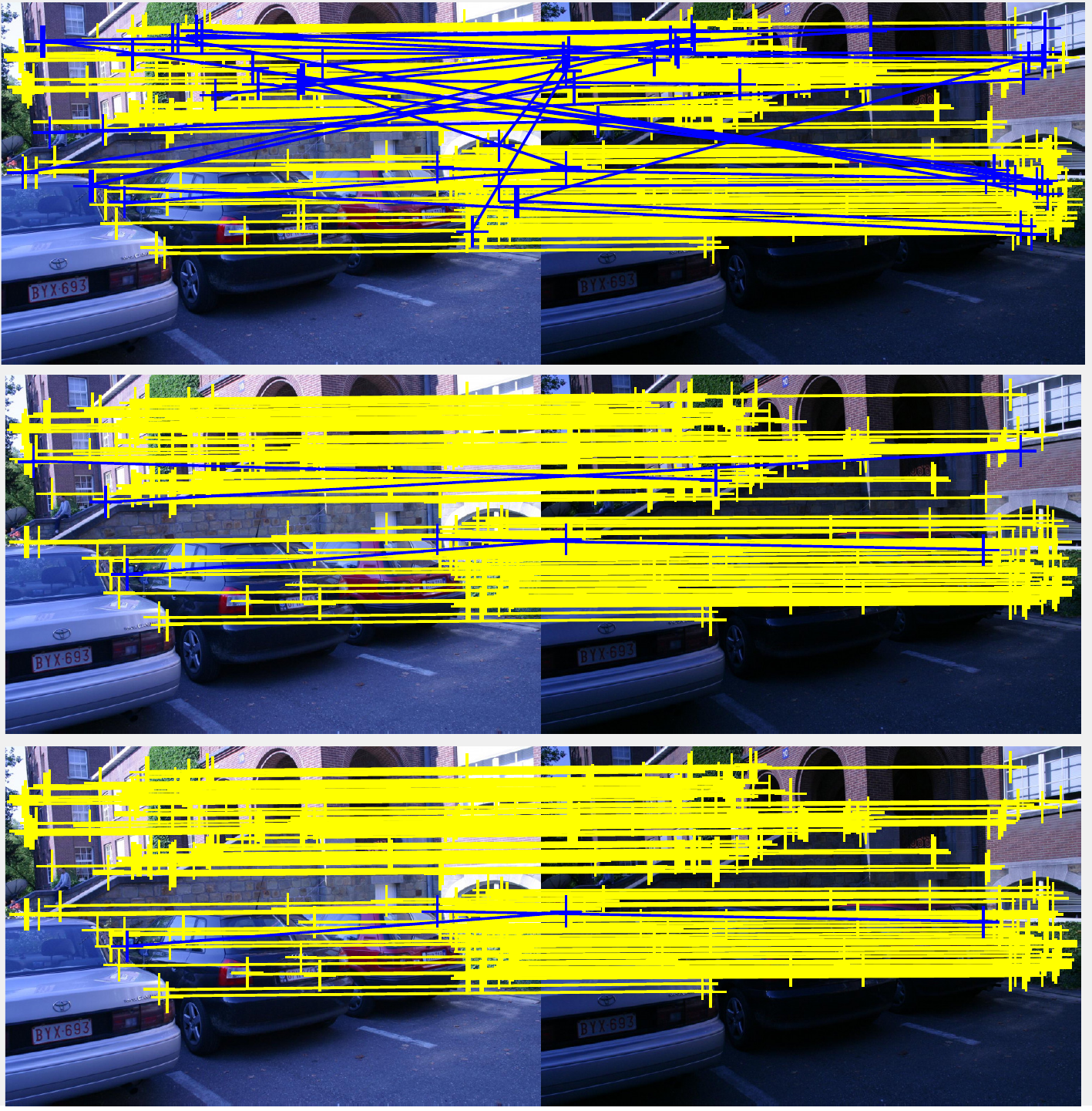}
\caption{Leuven}
\end{subfigure}
\caption{Example of matching results. The bottom one is from DMatch, the middle one is from MatchALS, and the top one is from original pairwise matching respectively. Yellow lines encode the correct matches, while blue lines are for wrong ones.}
\label{fig:matchings}
\end{figure*}

\subsubsection{Graffiti datasets}
In this experiment, we followed the procedure described in \cite{Zhou2015}. We used the benchmark datasets from Graffiti datasets \footnote{\url{http://www.robots.ox.ac.uk/~vgg/data/data-aff.html}}. In each dataset, there are 6 images of a scene with various image transformations, including viewpoint change, blurring, and illumination variation etc.

To construct an affinity score matrix $\bbx$, we employed the same procedure as in \cite{Zhou2015} for comparison purpose. We first detected 1000 affine covariant feature \cite{Miko2005} points in each image of the dataset and computed their SIFT \cite{lowe2004} descriptors using VLFeat library \cite{vlfeat}. The affinity scores were computed as the inner product between every pair of detected feature points on each pair of images.
%In our experiment, we found that the algorithm favors sparse set of affinity score, which have a better convergence property. Hence,
We excluded apparent mismatches by keeping only affinity scores that are above the threshold $0.7$. Furthermore, any potential matches that are indistinguishable was removed, i.e. if the first and the second top matches were below the ratio threshold $1.1$, the candidate point was removed. Finally, any feature point that has only one candidate match %in all other images
in the dataset was also excluded.

In a comparison, to construct our cover graph, % from the pairwise matching scores. Given all the pairwise matching scores,
we first built a matching quality graph, using the matching score as the edge weight and %find the laplacian embedding of the graph to cluster the images into two parts, i.e. we
used the Fiedler vector of the graph laplacian as the embedding and applied Algorithm \ref{alg:greedyK} to build the cover complex.
%to split the images. In order to have an overlap between the two cover nodes, we include a buffer at the clustering boundary to have one image on the positive side fuse into the negative cluster and vice versa. This way each cover node contains 4 images and the overlap between them has two images.

To evaluate the performance, we used the ground truth homography matrix given in the dataset, and adopted the procedure used in \cite{Chen2014}. For a testing point, we calculated the true correspondence using homography and compared with the matched correspondence. If they were within a predefined distance threshold, we deemed the matching is correct, and otherwise, wrong. Then we swept along the threshold dimension to draw an error curve that is dependent on the threshold chosen.

We tested our DMatch algorithm against MatchALS~\cite{Zhou2015}, MatchLift~\cite{Chen2014}, Spectral~\cite{Pachauri2013} and the original pairwise matchings. We ignored the Consensus algorithm \cite{Leonardox2017} as it cannot explicitly handle partial matches. %In \cite{Zhou2015}, a fair comparison has been done using the same dataset with MatchLift \cite{Huang2013} and Spectral method \cite{Pachauri2013}. It has been shown that MatchALS performed similarly with MatchLift and both outperformed Spectral method. To avoid overloading the figure, we hence omit the curves for MatchLift and Spectral method.
Figure~\ref{fig:error_curve} shows the curve for three datasets, Graffiti, Bikes, and Leuven. Note that DMatch will not give a full pairwise matching between images, instead, we only have a matching when the two images belong to the same cover node. Therefore, we computed the one-hop composite match between image pairs across different cover nodes\footnote{For each image $i$ and $j$ not in the same cover node, we loop through all $k\neq i,j$ and accumulate the composite matchings from $i\rightarrow k$ and $k\rightarrow j$.}. From the performance curve, we can see that our DMatch performs very similar to the best global methods in all datasets as shown in \cite{Zhou2015}. In another word, DMatch achieved performance gains without loss of matching quality.

In Figure \ref{fig:matchings}, we show the example matches between the first and the fourth image for each dataset. The bottom match is DMatch, the middle is MatchALS and the top one is the original pairwise map. Clearly, our matching shows at least as good as the results of MatchALS, where both corrected mismatches (reduced blue lines) and increased correct matches (denser yellow lines).

%It have been shown that our algorithm is completely comparable with MatchALS and both supersedes the performance of original pairwise matching.
In our implementation %\footnote{Thanks for the authors of \cite{Zhou2015} to kindly provide their original implementation of the algorithm.},
we notice that the total number of iterations to converge for both DMatch and MatchALS are roughly the same (around 60 iterations).
%Since we are implementing the algorithm on a single laptop, we are not able to provide any meaningful comparison on the timing of the two algorithms. Our intuition is, however, DMatch is much more scalable than MatchALS from the distributed nature of our algorithm.

\section{Conclusion}
In this paper, we introduced a scalable framework for establishing consistent matches across multiple graphs in a distributed manner.
%Our framework is applicable to a vwill allow us to extend any algorithm that use matching data matrix as representation format.
We showed how to use our framework to extend state-of-the-art global methods. By running an iterative optimization algorithm locally and exchange information in every iteration, our framework would achieve local and global consistent matching at the same time. Furthermore, we theoretically proved the sufficient conditions under which locally consistent matching would guarantee global consistency. In our experiments, we showed that in practice, the assumptions and the conditions in our theorem could be relaxed without sacrificing performance. In addition, our proposed distributed framework achieved order of magnitude improvements in speed. We believe this is a very important first step for large scale exploration of images for object matching as well as building 3D object models from crowd-sourced collections. Future work includes matching large collection of different deformable objects that have high similarity and enough variance, e.g. a collection of different dogs or cats.

\vspace{10pt}
\noindent{\bf Acknowledgements:} The authors wish to thank the support of NSF grants DMS-1546206 and DMS-1700234, a Google focused research award and a gift from Adobe Research.

{\small
\bibliographystyle{ieee}
\bibliography{hks}
}

\end{document}